\documentclass[oneside,12pt]{article}
\usepackage{eurosym}
\usepackage{amsmath}
\usepackage{amsfonts}
\usepackage{graphicx,epsfig}
\usepackage{epstopdf}
\usepackage{amssymb}

\begin{document}

\title{\textbf{\LARGE FLRW cosmological models with quark and strange quark matters in
f(R,T) gravity}}
\author{{\large Ritika Nagpal}$^{1}$ {\large , J. K. Singh}$^{2}$, {\large S. Ayg\"{u}n}$^{3}$ \\
$^{1,2}$\textit{Department of Mathematics,}\\
\textit{\ \ Netaji Subhas Institute of Technology,}\\
\textit{\ \ Faculty of Technology, University of Delhi,}\\
\textit{\ \ New Delhi-110 078, India}\\
$^{3}$\textit{Department of Physics,}\\
\textit{Arts and Sciences Faculty,}\\
\textit{ Canakkale Onsekiz Mart University,}\\
\textit{ Terzioglu Campus, 17020, Turkey}\\
\textit{\ \ ritikanagpal.math@gmail.com }$^{1}$, \textit{jainendrrakumar@rediffmail.com}$^{2}$,\\
\textit{saygun@comu.edu.tr}$^{3}$
}
\maketitle

{\footnotesize \vskip0.4in \noindent \textbf{Abstract} In this paper, we have studied the magnetized quark matter (QM) and strange quark matter (SQM) distributions in the presence of $ f(R,T)$ gravity in the background of Friedmann--Lema\^{\i}tre--Robertson--Walker (FLRW) metric. To get exact solutions of modified field equations we have used $f(R,T ) = R + 2 f(T)$ model given by Harko et al. with two different parametrization of geometrical parameters \textit{i.e.} the parametrization of the deceleration parameter $ q $, and the scale factor $ a $ in hybrid expansion form. Also, we have obtained Einstein Static Universe (ESU) solutions for QM and SQM distributions in $f(R,T)$ gravity and General Relativity (GR). All models in $f(R,T)$ gravity and GR for FRW and ESU Universes with QM also SQM distributions, we get zero magnetic field. These results agree with the solutions of Akta{\c s and Ayg\"{u}n in $f(R,T)$ gravity. However, we have also discussed the physical consequences of our obtained models.}

{\footnotesize \vskip0.2in \noindent \textbf{Keywords} }{\footnotesize \ $ f(R,T) $ theory, Magnetized Quark and Strange quark matter, FLRW universe, Deceleration parameter, Static Einstein universe}

\section{\protect\Large Introduction}

\noindent \qquad As we know cosmology, the study of Universe as a whole has been turned on its head by an impressive discovery that the Universe is flying apart in all the directions with an increasing rate. Recent observations of type Ia supernovae proclaimed that our Universe is not only expanding but also accelerating. Other observations \textit{e.g.} Cosmic Microwave Background Radiation (CMBR), Large-Scale Structure (LSS), Baryon Acoustic Oscillation (BAO) and Planck data \cite{per}-\cite{ade} also indirectly supports the accelerating expansion of the Universe.  However, some cosmologists believe on the decelerating model of the Universe with the dominant force gravity. To know the actual mechanism behind decelerating or accelerating expansion is still a point of discussion. Also, a big question the astronomers have to deal with the problem of initial singularity \textit{i.e.} the Universe must have emerged from the dense and hot primordial state (primeval atom) called \textit{big bang} and expanded to the present state. With enough matter outside, the dominant force, gravity collapse the whole Universe one day into itself called \textit{big crunch}. In 1930s Fritz Zwicky and later in the 1960-70s modern calculations were made by Vera Rubin, suggested that galaxies were spinning more quickly than they should, for which they suggested that there must be something beyond standard matter that holds the galaxies dubbed as \textit{Dark Matter} (DM). To explain the accelerating expansion of the Universe, a hypothetical form of energy is needed with high negative pressure dubbed as dark energy DE whose density is around $ 10^{-29} gm/cm^3 $ (near the critical density) which is approximately $ 5 $ hydrogen atoms in a $ cm^3 $. Recent Plank data suggested that the DM and DE constitute around $26\%$ and $ 69\%$ respectively and residue is normal matter. The simplest example of this puzzle takes us back to the Einstein's equation and the cosmological constant $ \Lambda $. Long after the Big-Bang, vacuum energy exerted enough pressure over extremely large scale to push the Universe out and as the Universe grew larger, more and more came into the existence, cause the expansion to accelerate. 

To expound the recent affair of cosmic acceleration, many alternative theories of gravity have been evolved beyond $\Lambda$CDM model. One of the modification is to modify the General Theory of Relativity (GR) in which the origin of DE is connected by the adjustment of gravity, as it is described by Riemannian geometry, which is torsion-less and also a number of gravitational theories have been analysed which depicts the torsion effects in the extension of GTR. Some of these theories are f(R)theory \cite{fer}, $ f(T) $ teleparallel theory \cite{ben}, where $ T $ is the torsion scalar, Brans-Dicke Cosmology \cite{bra}, Saez-Ballester theory \cite{saez}. Recently Harko et al. \cite{har} has proposed one of the most prospective and efficient version of the alternative theories of gravity named as $ f(R,T) $ gravity, where $R$ and $T$ are the scalar curvature and trace of the stress energy tensor respectively. A lot of remarkable works have been done in $f(R,T)$ theory of gravity \cite{sin1}-\cite{sin4}.
 
In our work, we consider quark matter (QM) and strange quark matter (SQM) in the influence of magnetic flux in the framework of homogeneous and isotropic FRW Universe model with cosmological constant $\Lambda$ in $f(R,T)$ theory
of gravity. Because, one of the most important and interesting topics in cosmology and astrophysics is to observe, simulation, and investigate primitive magnetic fields \cite{Subramanian}. We are aware of the presence of a magnetic field in Neutron stars, Pulsars, our Milky Way galaxy (approximately $10^{-6}$ G) and in other galaxies. But we do not know how the magnetic field evolved in the Universe \cite{akt},\cite{Marinacci}. That's why this is a cosmological problem to solve and it is a good source of motivation to investigate magnetic fields \cite{Grasso}. However, magnetic fields may have affected a number of relevant continuum in the early Universe and the geometry of Universe itself \cite{Grasso}. Cosmic magnetic fields are known to play an important role in the formation of galaxies  \cite{Pebless},\cite{Wasserman}. Magnetic fields influence the elements that make up matter. The idea of a primitive magnetic field is quite attractive because it has the potential to explain large-scale fields in the Universe. Magnetic fields are a common and important component of the Universe \cite{akt},\cite{Carilli},\cite{Wolfe}. Therefore, it is very important in terms of information about structure formation in the early phase of Universe \cite{Grasso}, \cite{gou}. Lately, a number of studies have been carried out to clarify primitive magnetic fields and quark gluon matter in the early Universe.

Also, to understand the early phase of Universe, it is influential to study quark gluon plasma. Quarks are the most fundamental particles known as the building blocks of the matter we see around us. The term fundamental particle means the particle having no substructure that can be split further. After the Big-Bang at a critical temperature $T_c \equiv 100-200$ $MeV$ when a quark gluon phase transition of the Universe had taken place, the QM is thought to be emerged out. It is difficult to isolate a single quark, the reason behind this is that quarks never exist by themselves, they always exist in groups. Quarks are basically of 6 types: up(u), down(d), strange(s), charm(c), top(t) and bottom(b) in which up(u), down(d) and strange(s) are three main types of quarks. Nucleons, we know are mostly made of up and down quarks but in nuclei of strange matter, strange quark is also present and they have very high density. There is always a fractional charge associated with quarks like a up(u) quark has a charge of $+\frac{2}{3}$ whereas a down(d) type has $-\frac{1}{3}$ charge, for example we can say a neutron is made up of two down(d) and one up(u) type quark and proton is made of two up(u) and one down(d) type quarks. Strange quark has a charge of $-\frac{1}{3}$ and has its unique property called strangeness with value -1 whereas an anti strange quark has strangeness of +1.

According to various authors, SQM could be the true basis state of hadronic matter \cite{Bodmer}-\cite{Geng}. Also, according to the Bodmer-Witten hypothesis, the presence of strange quark stars (SSs) has been estimated based on the SQM hypothesis \cite{witten}. SSs may be naked SQM objects wrapped in simple nuclear crusts or bulk SQM nuclei \cite{weber}. For this reason, it is important to investigate the properties of SSs crusts, because the crust may describe the observational effects of SSs \cite{Hua}. The possible presence of nuclear crusts creates SSs that are very similar to normal neutron stars for a distant observer. In the framework of the SQM hypothesis, SQ dwarfs and even SQ planets may exist in a stable manner \cite{Huang}. Recently, Huang et al. \cite{Hua}, \cite{Huang}  have searched SQ planets and have investigated for detailed and important research on the possibility of identifying SQM.

However, the equation of state (EoS) of SQM is given by $p=\frac{\rho-4B_c}{3}$, where $B_c$ is the Bag constant that the difference between the energy density of the perturbed and non-perturbed QCD vacuum established on the Bag model of QM, $p$ is the thermodynamics pressure and $\rho$ is the energy density of the QM. In the simplest adaptation of the Bag model, one can assume that the quarks are mass-less and non-interactive and consider the EoS of the form $ p_q=\frac{\rho_q}{3} $. The total pressure and energy density are $ p=p_q-B_c $ and $ \rho = \rho_q+B_c $ respectively. The number of studies have been done by considering QM and SQM with GR and other modified theories of gravity  \cite{cagl}-\cite{kat}. Recently, P. K. Sahoo \textit{et al}. \cite{sah1},\cite{sah2} have constructed an anisotropic models with magnetized SQM in $ f(R,T) $ gravity by considering some specific parametrization of deceleration parameter.

Following \cite{sah1}, in this paper, we have studied flat FLRW model with magnetized QM and SQM in $ f(R,T) $ gravity by considering two different parametrization of geometrical parameters \textit{i.e.} deceleration parameter $ q $ and scale factor $ a $. In the first parametrization, we have considered the form of deceleration parameter proposed by \cite{abd} which is a bouncing model, and in the second parametrization, a form of scale factor is discussed by \cite{ozg},\cite{mish} known as hybrid scale factor (HSF) cosmology. The work is organized in various sections as follows: Sect. 1 is introductory. The general Einstein-Hilbert action on the $ f(R,T) $ gravity \cite{har} with essential literature review along with metric and field equations is discussed in Sect. 2. The dynamics of the Universe for the varying deceleration parameter and HSF cosmology is discussed in Sect. 3. The solutions of the EFE under the influence of $ f(R,T) $ gravity have been derived in Sect. 4 for Model I, II and ESU Model. In Sect. 5, we have discussed the physical interpretations and validation of the models through various physical consequences. Finally, we have summarized and concluded our work for Model-I and II in the last Sect. 6.

\section{The background metric and field equations}

\qquad The $f(R,T)$ gravity \cite{har} which is a modified theory and depending upon the coupling between geometry of the Universe and the matter present in the Universe, is the extension of $f(R)$ gravity or more precisely GR. The formalism of $f(R,T)$ model depends on a source term which is a function of matter Lagrangian $ L_m $. The action of $f(R,T)$ gravity is given by
\begin{equation}
S=\frac{1}{16\pi G}\int \sqrt{-g}[f(R,T)+L_{m}]d^{4}x,  \label{1}
\end{equation}
where $g$ is the determinant of the metric $g_{\mu \nu}$. On varying the action given in (\ref{1}) \textit{w.r.t.} $g_{\mu \nu}$, we have

\begin{equation}
G_{\mu \nu }+\left( g_{\mu \nu }\square -\nabla _{\mu }\nabla _{\nu }\right)
=[8\pi +2f^{\prime }(T)]T_{\mu \nu }+2[f^{\prime }(T)p+\frac{1}{2}f(T)]g_{\mu \nu }+\Lambda g_{\mu \nu},  \label{2}
\end{equation}
for which we consider the $f(R,T)=R+2f(T)$, combination of a scalar and tensor quantity namely, Ricci scalar $R$ and stress-energy tensor $T$. The prime denotes derivative \textit{w.r.t.} trace T. We consider $f(T)=\lambda T$, $\lambda$ being a coupling constant. In different forms of $f(R,T)$ function given by Harko \cite{har}, we assume here an elementary form in this formalism through which GR can be restored by putting $\lambda=0$. We choose matter Lagrangian $L_{m}=-p_{m}$ in the action (\ref{1}).\\

We consider the background metric in the form of homogeneous, isotropic and spatially  Robertson-Walker geometry given by 
\begin{equation} \label{3}
ds^{2}=dt^{2}-a^{2}\left[ \frac{dr^2}{1-\kappa r^2}+r^2(d\theta^2+\sin^2 \theta d\phi^2)   \right], 
\end{equation}
where the scale factor $a$ is a function of cosmic time $t$ and $\kappa=0, \pm 1$. The matter in the Universe as the source of gravitation consists of magnetized QM, SQM which are involved in the energy momentum tensor

\begin{equation}\label{4}
T_{\mu \nu }=\big(\rho +p+h^{2}\big)u_{\mu }u_{\nu }-\Big(\frac{h^{2}}{2}+p%
\Big)g_{\mu \nu }-h_{\mu }h_{\nu },  
\end{equation}
where $ \rho $, $ p $ and $ h^{2} $ are the energy density, isotropic pressure and magnetic flux respectively \cite{Tsa, Ay}. Here, the magnetic flux $ h^{2} $ is considered in the $ x $-direction as $ h_{i}u^{i}=0 $.

In the cosmological background of $f(R,T)$ theory, the Einstein field equations (\ref{2}) with cosmological constant $\Lambda $ takes the form
\begin{equation}\label{5}
G_{\mu \nu }=\big(8\pi +2f^{\prime }(T)\big)T_{\mu \nu }+\big(2f^{\prime
}(T)p+f(T)+\Lambda \big)g_{\mu \nu },  
\end{equation}%
Here $T$, the trace of the stress energy momentum tensor is given by $ T=\rho-3p$. So the Eq. (\ref{5}) becomes 
\begin{equation} \label{6}
G_{\mu \nu }=\big(8\pi +2\lambda \big)T_{\mu \nu }+\big(\lambda \rho -\lambda p+\Lambda \big)g_{\mu \nu }. 
\end{equation}\\
The sequence of gravitational field equations are given as
\begin{equation}\label{7}
\frac{2\ddot a}{a}+\frac{\dot a^2}{a^2}+\frac{\kappa}{a^2}=(4\pi +\lambda )h^{2}-(8\pi +3\lambda )p+\lambda \rho +\Lambda ,
\end{equation}
\begin{equation}\label{8}
\frac{2\ddot a}{a}+\frac{\dot a^2}{a^2}+\frac{\kappa}{a^2}=-(4\pi +\lambda )h^{2}-(8\pi +3\lambda )p+\lambda \rho +\Lambda,
\end{equation}
\begin{equation} \label{9}
\frac{3\dot a^2}{a^2}+\frac{3\kappa}{a^2}=(4\pi +\lambda )h^{2}+(8\pi +3\lambda )\rho -\lambda p+\Lambda . 
\end{equation}\\
In order to find physically viable solutions of models, we assume the equation of state (EoS) for QM for SQM as
\begin{equation}\label{10}
p_{q}=\omega \rho _{q},\,\,\,0\leq \omega \leq 1,  
\end{equation}
and 
\begin{equation}\label{11}
p=\omega (\rho -\rho _{\ast }),  
\end{equation}
respectively, where $ \rho _{\ast } $ indicates the energy density when pressure $ p $ is zero. If we take $\omega =\frac{1}{3}$ and $ \rho _{\ast }=4B_{c} $, then Eq. (\ref{11}) transformed into the EoS of SQM in Bag model \cite
{kap,sot} 
\begin{equation} \label{12}
p=\frac{\rho -4B_{c}}{3}, 
\end{equation}%
where $B_{c}$ denotes the Bag constant. To get an explicit solution of Einstein field equations (EFE) for QM and SQM in $ f(R,T) $ gravity, we need one more constrain equation for the consistency of the above system of equations. In literature, there are several schemes of parametrization of cosmological parameters to complete the system for an explicit solution. Although, this is an adhoc choice but it do not affect the validity of background theory and can be a relevant way to study dark energy models. Here in this
paper, we consider a specific form of variable deceleration parameter in the first case and the hybrid scale factor (HSF) cosmology in second case. 

\section{Dynamics of the Universe for two parametrization schemes}

\qquad In our first assumption, we consider a specific form of deceleration parameter leading to a non-singular bouncing model which is of the form \cite{abd} 
\begin{equation} \label{13}
q=\frac{-\alpha }{t^{2}}+(\beta -1), 
\end{equation}
where $\alpha>0$ is a model parameter having dimension of square of time and $ \beta >1 $ is a dimensionless model parameter which are responsible for the dynamics of the Universe. As the deceleration parameter depends on the
cosmic time $t$, so $q$ may be positive or negative for a given set of constraints on $\alpha $ and $\beta $. In general, the model shows 
\begin{itemize}
\item[(1)]  decelerating expansion of the universe ($q>0$) when $\alpha <0$ and $%
\beta >1$ or $0<\alpha <(\beta -1)t^{2}$ and $\beta >1$,\newline
\item[(2)] accelerating expansion ($q<0$) when $\alpha >(\beta -1)t^{2}$ and $\beta >1$.
\end{itemize}
The Hubble parameter $ H $ is given by relation 
\begin{equation}\label{14}
q=-1+\frac{d}{dt}\Big(\frac{1}{H}\Big).  
\end{equation}\\
Integrating (\ref{14}), we get scale factor 
\begin{equation} \label{15}
a(t)=e^{\gamma }exp\int \frac{1}{\int \Big(\frac{-\alpha }{t^{2}}+\beta \Big)dt}dt+\delta , 
\end{equation}
where $\gamma $ and $\delta $ are arbitrary integrating constants. Solving (\ref{15}), we get an explicit determination of scale factor $a(t)$ of the form 
\begin{equation}
a(t)=e^{\gamma }exp{\int \frac{t}{\beta t^{2}+\eta t+\alpha }dt},
\label{16}
\end{equation}%
where $\eta $ is an arbitrary constant. Eq. (\ref{16}) have three different forms of scale factor $a(t)$ depends on the cases when $ \eta \neq 2\sqrt{\alpha \beta } $, $ \eta =2\sqrt{\alpha \beta } $,
and  $ \eta =0 $. Here, we consider the case when $ \eta =0 $ for which Eq. (\ref{16}) yields 
\begin{equation}
a(t)=e^{\gamma }exp\int \frac{t}{\beta t^{2}+\alpha }dt.  \label{17}
\end{equation}
Now making the calculation accessible and without the loss of generality, we take the integrating constant $\gamma =0$ for which the scale factor $a(t)$
takes the form 
\begin{equation}
a(t)=(\alpha +\beta t^{2})^{\frac{1}{2\beta }}.  \label{18}
\end{equation}\\
Using Eq. (\ref{18}), the functional form of Hubble parameter $ H $ for this model is obtained as 
\begin{equation}
H(t)=\frac{t}{\alpha +\beta t^{2}}.  \label{19}
\end{equation}
\qquad In our second assumption, we consider the parametrization as the hybrid scale factor cosmology  \cite{ozg,mish} in the form

\begin{equation}
a(t)=e^{k_{1}t}t^{k_{2}},  \label{20}
\end{equation}%
where the parameters $k_{1}$ and $k_{2}$ are non-negative and dimensionless. Power law cosmology can be obtained when $ k_{1}=0 $ and de-sitter Universe can be brought back when $ k_{2}=0 $. The cosmic dynamics in
power law cosmology $(t^{k_{2}})$ is dominated over the exponential term $ (e^{k_{1}t}) $ in the early phases of evolution the Universe and as the Universe evolves, the cosmic dynamics in the de-sitter Universe (\textit{i.e.}
exponential term $ e^{k_{1}t} $) is dominated over the second term $ (t^{k_{2}}) $ at late times. The time dependent deceleration parameter $q(t)$ and the Hubble parameter $H(t)$ of Hybrid scale factor takes the form

\begin{equation}
q(t)=-1+\frac{k_{2}}{(k_{1}t+k_{2})^{2}},  \label{21}
\end{equation}

\begin{equation} \label{22}
H(t)= k_1+\frac{k_2}{t}.
\end{equation}
Also a phase transition at $ t=\frac{-k_{2}}{k_{1}}+\frac{\sqrt{k_{2}}}{k_{1}} $ from early deceleration ($ q>0 $) to late time acceleration ($ q<0 $) of the Universe can be characterized by the Hybrid exponential law on constraining the parameters $ k_{1}>0 $ and $ 0<k_{2}<1 $. \\
{\scriptsize 
\begin{center}
{\scriptsize \textbf{Table 1.} The behavior of geometrical parameters for Model-I}\vskip0.1in 
\begin{tabular}{|c|c|c|c|}
\hline
{\small Model I} & ${\small q}$ & ${\small a}$ & ${\small H}$ \\ \hline
${\small t\rightarrow 0}$ & ${\small -\infty }$ & ${\small \alpha }^{\frac{1%
}{2\beta }}$ & ${\small 0}$ \\ \hline
${\small t\rightarrow \infty }$ & ${\small \beta -1}$ & ${\small \infty }$ & 
${\small 0}$ \\ \hline
\multicolumn{4}{|c|}{{\small Phase transition time }$t_{ph}=\sqrt{\frac{%
\alpha }{\beta -1}}$} \\ 
\hline
\end{tabular}
\end{center}
\begin{center}
{\scriptsize \textbf{Table 2.} The behavior of geometrical parameters for Model-II}\vskip0.1in
\begin{tabular}{|c|c|c|c|}
\hline
{\small Model II} & ${\small a}$ & ${\small q}$ & ${\small H}$ \\ \hline
${\small t\rightarrow 0}$ & ${\small 0}$ & ${\small -1+}\frac{1}{k_{2}}$ & $%
{\small \infty }$ \\ \hline
${\small t\rightarrow \infty }$ & ${\small \infty }$ & ${\small -1}$ & $%
{\small k}_{1}$ \\ \hline
\multicolumn{4}{|c|}{{\small Phase transition time }$t_{ph}=\frac{-k_{2}}{%
k_{1}}+\frac{\sqrt{k_{2}}}{k_{1}}$} \\ \hline
\end{tabular}
\end{center}
}
In Model I, the deceleration parameter $q\rightarrow -\infty $ as $t\rightarrow 0$ initially, reduces to zero when $ t=\sqrt{\frac{\alpha }{\beta -1}} $ \textit{i.e.} at the time of phase transition, and eventually takes a positive value $\beta-1$, $ \beta>1  $ as $ t \to \infty $. From Eq. (\ref{19}), it is easy to observe that the initial time  $ t=0 $ is the bouncing point since $ H=0 $, $ a(t)=\alpha ^{\frac{1}{2\beta }}=a_{0} $(say) $ \neq 0 $ but $\dot{a}=0$,  and $\ddot{a}=\frac{a_{0}}{\alpha }=constant$, which obviously shows that our model is free from initial singularity and begins with a finite acceleration. Also we find that $H(t)\rightarrow 0$ as $t\rightarrow 0, \infty$. In Model II, initially $ a(t)\to 0 $ as $ t \to 0 $, which shows that the Universe evolves with zero volume and attains infinite volume at late time, which shows the standard Big-Bang scenario. Also this model represents the decelerating phase ($ q>0 $) at the early Universe, phase transition ($ q=0 $) at $ t=\frac{-k_{2}}{k_{1}}+\frac{\sqrt{k_{2}}}{k_{1}} $, and then get entered into the accelerating phase ($ q<0 $) at late time. 

\section{Exact solution}
\subsection{Model-I}

\qquad The field equations (\ref{7})-(\ref{9}) contain four unknowns $ \rho ,\,p,\,\Lambda ,\,h^{2} $. In order to get physically viable solutions, we assume EoS parameters (\ref{10}), (\ref{11}) for QM and SQM respectively. Using Eqs. (\ref{7})-(\ref{10}) and (\ref{18}), we obtain the value of magnetic flux $ h^{2} $, energy density $ \rho _{q} $, isotropic pressure $ p_{q} $ and cosmological constant $ \Lambda_{q} $ for QM as follows: 
\begin{equation} \label{23}
h^{2}=0,
\end{equation}
\begin{equation} \label{24}
\rho _{q}=\frac{\beta t^{2}-\alpha}{(4\pi +\lambda )(1+\omega )(\alpha +\beta t^{2})^{2}}+\frac{\kappa}{(4\pi+\lambda)(1+\omega )(\alpha +\beta t^{2})^{\frac{1}{\beta}}},
\end{equation}
\begin{equation}\label{25}
p_{q}=\frac{\omega(\beta t^{2}-\alpha)}{(4\pi +\lambda )(1+\omega )(\alpha +\beta t^{2})^{2}}+\frac{\kappa \omega}{(4\pi+\lambda)(1+\omega )(\alpha +\beta t^{2})^{\frac{1}{\beta}}},
\end{equation}
\begin{equation}\label{26}
\Lambda _{q}=\frac{[((\beta+3)\omega-3\beta+3)t^2-\alpha(\omega-3)]\lambda}{(4\pi +\lambda )(1+\omega )(\alpha +\beta t^{2})^{2}}+\frac{4\pi[(3\omega-2\beta+3)t^2+2\alpha]}{(4\pi +\lambda )(1+\omega )(\alpha +\beta t^{2})^{2}}+\frac{4(3\pi \omega+\lambda \omega +\pi)\kappa}{(4\pi+\lambda)(1+\omega )(\alpha +\beta 	t^{2})^{\frac{1}{\beta}}}.
\end{equation}

Using Eqs. (\ref{7})-(\ref{9}), (\ref{12}) and (\ref{18}), we have the value of magnetic flux $ h^{2} $, energy density $ \rho _{sq} $, isotropic pressure $ p_{sq} $, EoS parameter $ \omega_{sq} $ and cosmological constant $\Lambda_{sq} $ for SQM as follows: 
\begin{eqnarray}  \label{27}
h^2=0,
\end{eqnarray}

\begin{eqnarray}  \label{28}
\rho_{sq}=\frac{3(\beta t^2-\alpha)}{4(4\pi+\lambda)(\alpha+\beta t^2)^2}+\frac{3\kappa}{4(4\pi+\lambda)(\alpha +\beta t^{2})^{\frac{1}{\beta}}}+B_c,
\end{eqnarray}

\begin{eqnarray}  \label{29}
p_{sq}= \frac{\beta t^2-\alpha}{4(4\pi+\lambda)(\alpha+\beta t^2)^2}+\frac{\kappa}{4(4\pi+\lambda)(\alpha +\beta t^{2})^{\frac{1}{\beta}}}-B_c,
\end{eqnarray}

\begin{eqnarray}  \label{31}
\Lambda_{sq}= \frac{(2\alpha-(2\beta-3)t^2)\lambda}{(4\pi+\lambda)(\alpha+\beta t^2)^2}-\frac{6\pi((\beta-2)t^2-\alpha)}{(4\pi+\lambda)(\alpha+\beta t^2)^2}+\frac{(6\pi+\lambda)\kappa}{(4\pi+\lambda)(\alpha +\beta t^{2})^{\frac{1}{\beta}}}-4(2\pi+\lambda)B_c.
\end{eqnarray}

Here, we are interesting to study the accelerating Universe only, therefore we consider the constraint $\alpha >(\beta -1)t^{2}$ and $\beta >1$ in Model I. We choose $\alpha=2(\beta-1)t^2$ to satisfy this constraint.
The graphical representations of energy densities ($ \rho_q, \rho_{sq} $), isotropic pressure $ p_{sq} $, EoS parameter $\omega_{sq} $ and the cosmological constants ($ \Lambda_q, \Lambda_{sq} $) for Model I with the values \textit{$ \lambda=1 $, $ \beta =1.1 $} and \textit{$ B_c=60 $} for QM and SQM are shown in Fig. 1, 2 and 3 respectively.\\

\begin{figure}[tbh]
\begin{center}
$%
\begin{array}{c@{\hspace{.1in}}cc}
\includegraphics[width=2.8in]{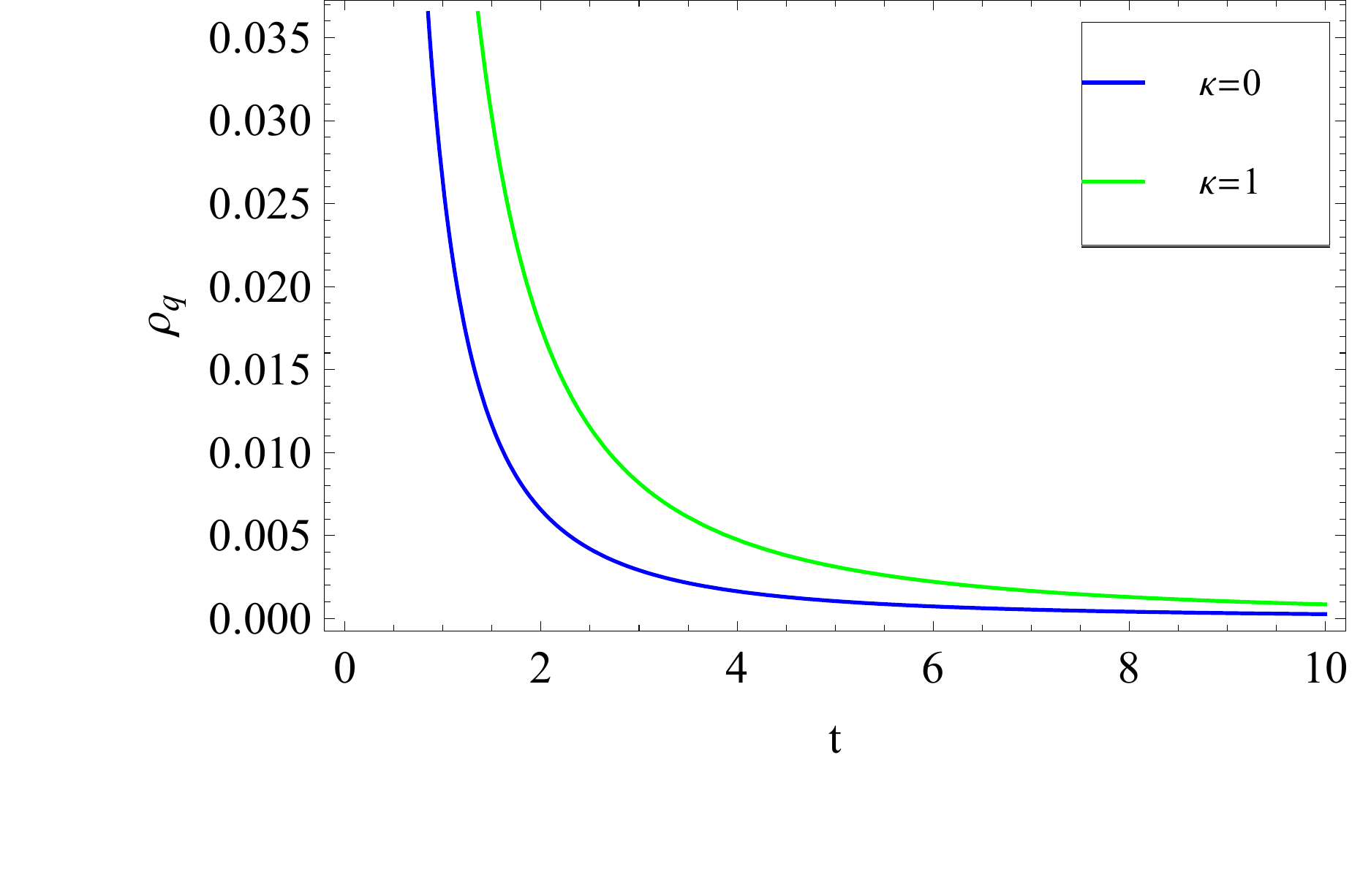} & 
\includegraphics[width=2.8in]{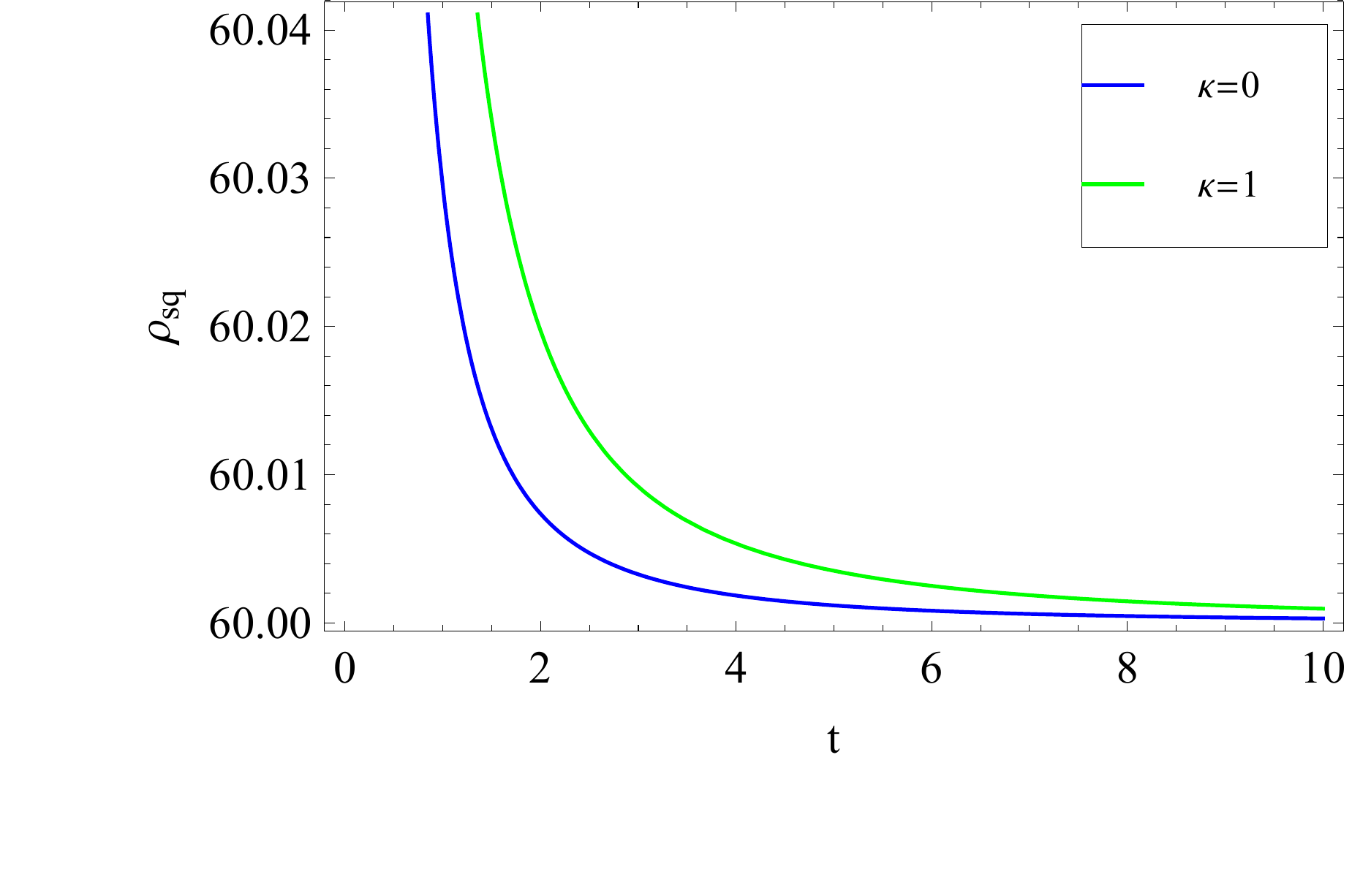}  \\ 
\mbox (a) & \mbox (b)%
\end{array}%
$%
\end{center}
\caption{\scriptsize (a) \textit{ The plot of $ \rho_{q} $  Vs. $ t $ for QM for Model-I }.\,\,\,  (b) \textit{ The plot of $ \rho_{sq} $  Vs. $ t $ for SQM for Model-I. }}
\end{figure}

\begin{figure}[tbh]
\begin{center}
$%
\begin{array}{c@{\hspace{.1in}}cc}
\includegraphics[width=2.8in]{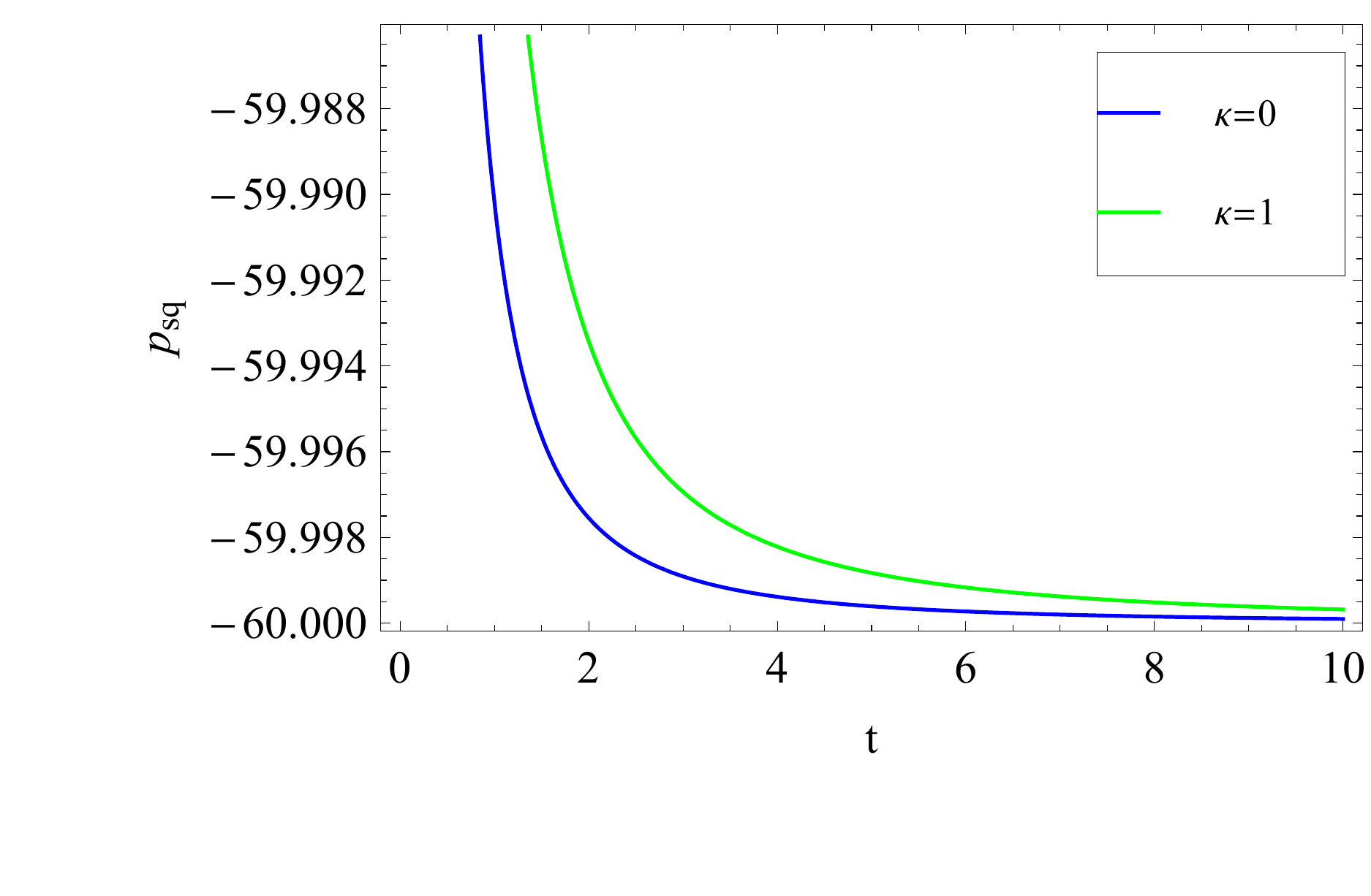} & 
\includegraphics[width=2.8in]{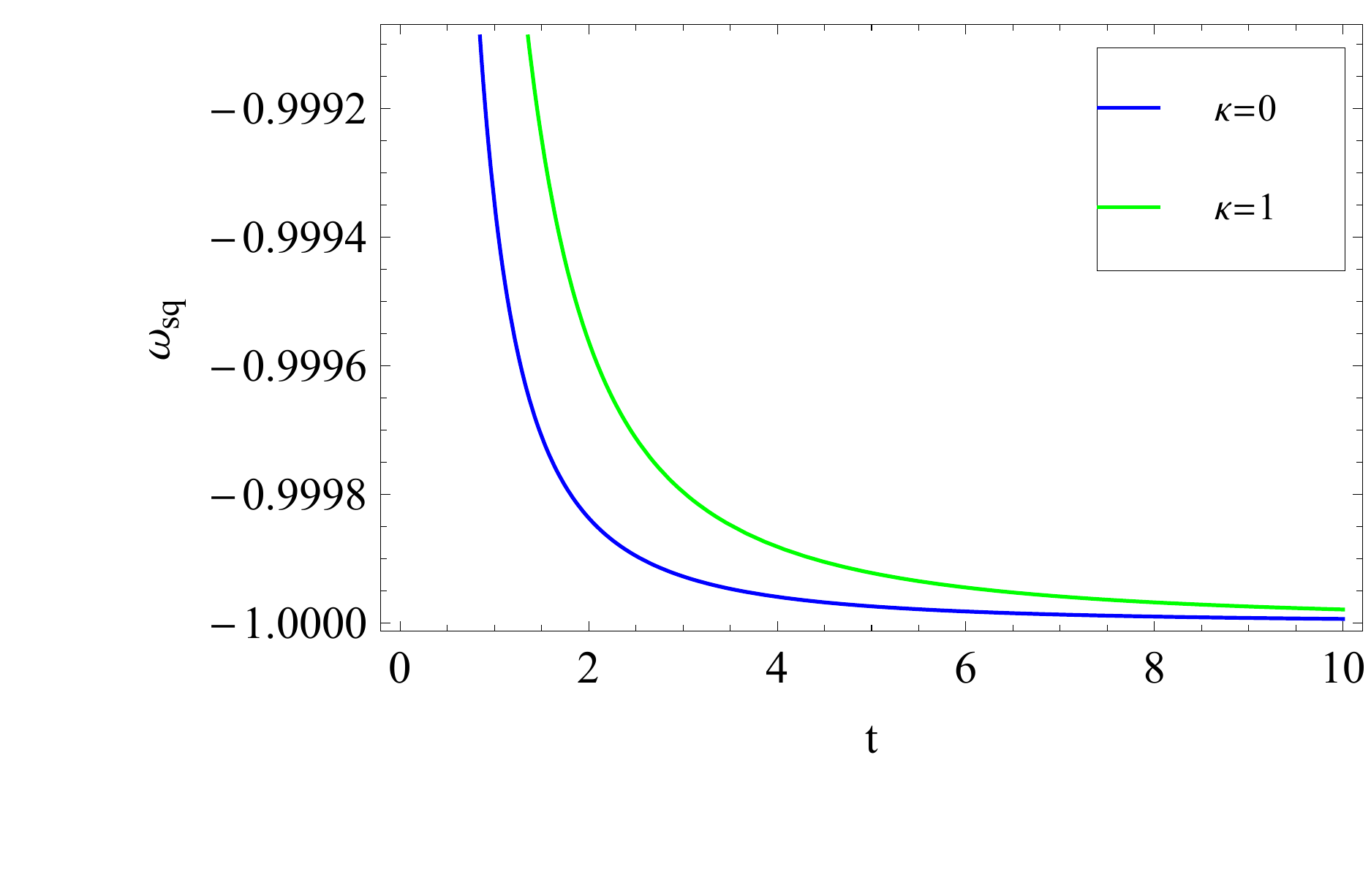}  \\ 
\mbox (a) & \mbox (b)%
\end{array}%
$%
\end{center}
\caption{\scriptsize (a) \textit{ The plot of $ p_{sq} $  Vs. $ t $ for SQM for Model-I }.\,\,\,  (b) \textit{ The plot of $ \omega_{sq} $  Vs. $ t $ for SQM for Model-I. }}
\end{figure}

\begin{figure}[tbh]
\begin{center}
$%
\begin{array}{c@{\hspace{.1in}}cc}
\includegraphics[width=2.8in]{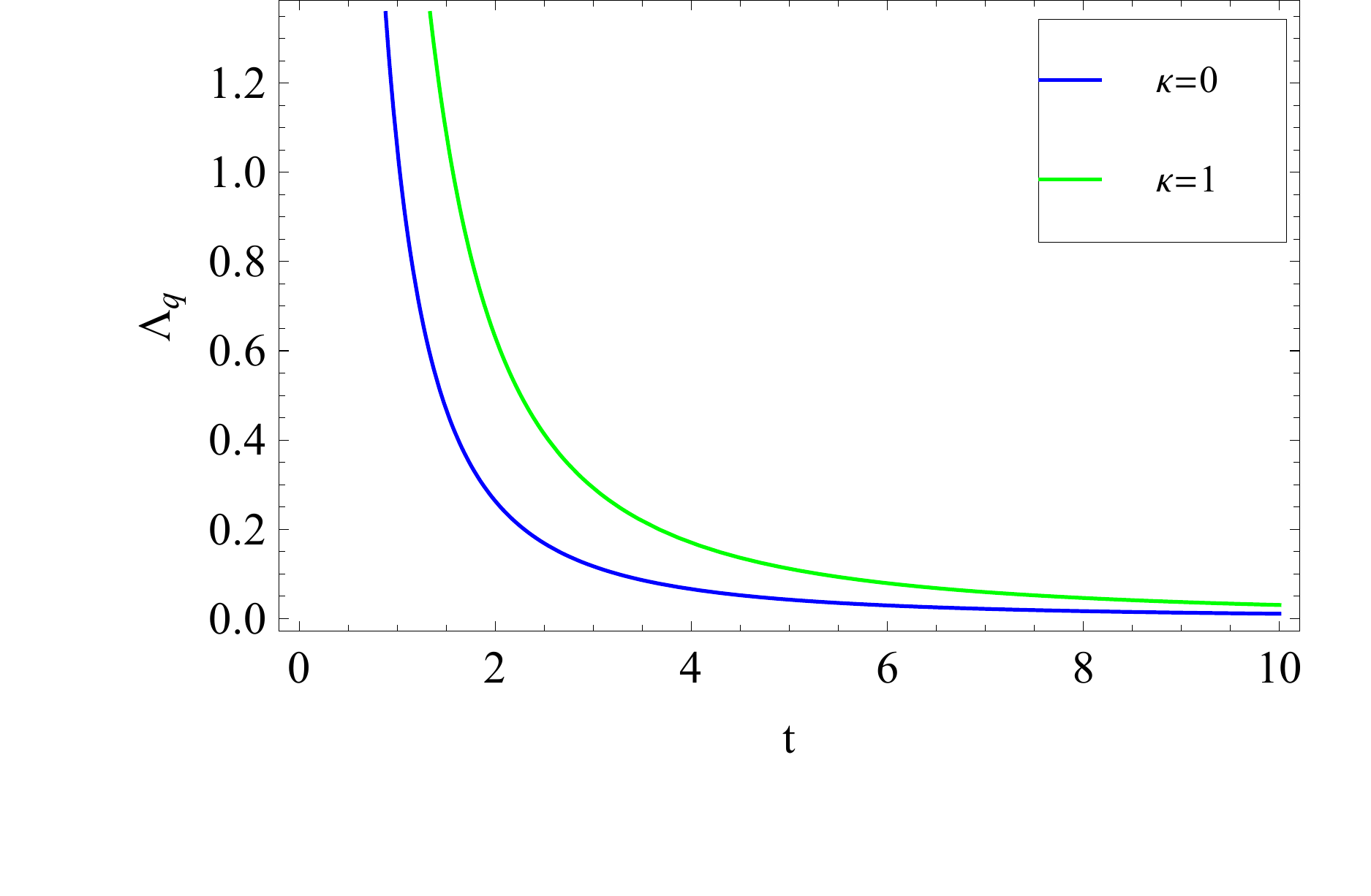} & 
\includegraphics[width=2.8in]{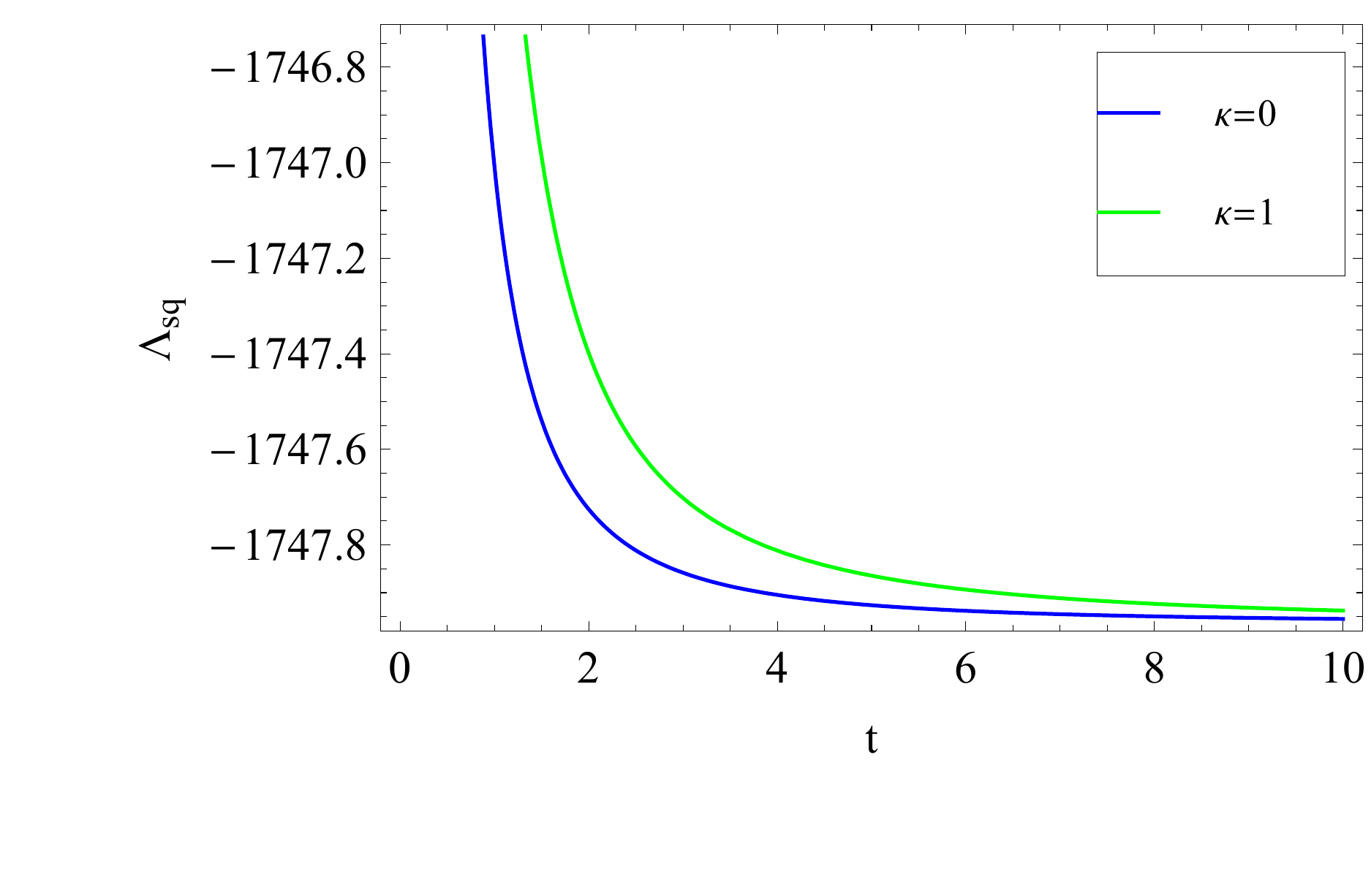}  \\ 
\mbox (a) & \mbox (b)%
\end{array}%
$%
\end{center}
\caption{\scriptsize (a) \textit{ The plot of $ \Lambda_{q} $  Vs. $ t $ for QM for Model-I }.\,\,\,  (b) \textit{ The plot of $ \Lambda_{sq} $  Vs. $ t $ for SQM for Model-I. }}
\end{figure}

Fig. 1(a) shows the profile of energy density $ \rho_q $ for QM in Model I for flat and closed Universe. Initially at the time of  evolution of the Universe, the amount of $ \rho_q $ is very large for both $\kappa=0,1$ which steadily reduces with time and $ \rho_q \to 0 $ as $ t \to \infty $. Model has maximum value of energy density for both the Universe ($\kappa=0,1$) near $t=0$.  Fig. 1(b) represents the behavior of energy density $\rho_{sq}$ for SQM in Model I for $\kappa=0,1$. It has been observed readily that $\rho_{sq}$ behaves alike $\rho_q$ with only one diversity is that $\rho_{sq} \to B_c$ as $t \to \infty$.\\

Fig. 2 demonstrates the profile of pressure $ p_{sq} $ and $ \omega_{sq} $ for SQM. Fig. 2(a) describes the pressure profile $ p_{sq} $ of Model I which is a decreasing function of time, remains negative throughout the evolution and $ p_{sq}\rightarrow -B_{c} $ as $ t\rightarrow \infty $ for Model I. The negative pressure $ p_{sq} $ corresponds to the accelerating expansion of the Universe. Fig. 2(b) exhibits the behavior of EoS parameter $ \omega_{sq} $ for SQM in Model I. The EoS parameter $ \omega_{sq} $ belongs to the quintessence domain, decreases with time and finally approaches to phantom divide line \textit{i.e.} $ \omega_{sq} \to -1 $ in late time which is affirmative with the current observations. Thus, in this case, we see that SQM behaves like dark energy.\\ 

Fig. 3(a) illustrates the behavior of cosmological constant $\Lambda_q$ for QM. In Model I, $\Lambda_q$ is monotonically decreasing function of time and $\Lambda_q \to 0$ as $t \to \infty$. 
Fig. 3(b) highlights the action of cosmological constant $\Lambda_{sq}$  for SQM. In Model I, $\Lambda_{sq}$ decreases with time and the graph of $\Lambda_{sq}$ remains negative throughout.\\

\subsection{Model-II}
\qquad Using Eqs. (\ref{7})-(\ref{10}) and (\ref{20}), we get the magnetic flux $ h^2 $, energy density $ \rho_q $, isotropic pressure $ p_q $ and cosmological constant $ \Lambda_q $ for QM as follows: 
\begin{eqnarray}  \label{32}
h^2=0,
\end{eqnarray}
\begin{eqnarray}  \label{33}
\rho_q = \frac{k_2}{(4\pi+\lambda)(1+\omega)t^2}+\frac{\kappa}{(4\pi+\lambda)(1+\omega)t^{2k_2}e^{2k_1t}}.
\end{eqnarray}

\begin{eqnarray}  \label{34}
p_q = \frac{k_2\omega}{(4\pi+\lambda)(1+\omega)t^2}+\frac{\kappa \omega}{(4\pi+\lambda)(1+\omega)t^{2k_2}e^{2k_1t}},
\end{eqnarray}
\begin{eqnarray}  \label{35}
\Lambda_q =\frac{k_2(\lambda (\omega-3)-8\pi)}{(4\pi+\lambda)(1+\omega)t^2}+3\left(k_1+\frac{k_2}{t}\right)^2 +\frac{4\kappa (3\pi \omega+\lambda \omega+\pi)}{(4\pi+\lambda)(1+\omega)t^{2k_2}e^{2k_1 t}}
\end{eqnarray}

Using Eqs. (\ref{7})-(\ref{9}), (\ref{12}) and (\ref{20}), we obtain the magnetic flux $ h^2 $, energy density $ \rho_{sq} $, isotropic pressure $ p_{sq} $, EoS parameter $ \omega_{sq} $ and cosmological constant $\Lambda_{sq} $ for SQM as follows: 

\begin{eqnarray}  \label{36}
h^2=0,
\end{eqnarray}

\begin{eqnarray}  \label{37}
\rho_{sq}=\frac{3k_2}{4(4\pi+\lambda)t^2}+\frac{3\kappa}{4(4\pi+\lambda)t^{2k_2}e^{2k_1t}}+B_c,
\end{eqnarray}

\begin{eqnarray}  \label{38}
p_{sq}= \frac{k_2}{4(4\pi+\lambda)t^2}+\frac{\kappa}{4(4\pi+\lambda)t^{2k_2}e^{2k_1t}}-B_c,
\end{eqnarray}

\begin{eqnarray}  \label{40}
\Lambda_{sq}=\frac{-2k_2(3\pi+\lambda)}{(4\pi+\lambda)t^2}+   3\left(k_1+\frac{k_2}{t}\right)^2+\frac{(6\pi+\lambda)\kappa}{(4\pi+\lambda)t^{2k_2}e^{2k_1t}}  -4(2\pi+\lambda)B_c.
\end{eqnarray}

The graphical representations of energy densities ($ \rho_q, \rho_{sq} $), isotropic pressure $ p_{sq} $, EoS parameter $\omega_{sq} $ and the cosmological constants ($ \Lambda_q, \Lambda_{sq} $) for Model II with the values \textit{$ \lambda=1 $, $ \omega=0.5 $, $ k_1=0.5 $, $ k_2 =0.6 $} and \textit{$ B_c=60 $} for QM and SQM are shown in Fig. 4, 5 and 6 respectively.

\begin{figure}[tbh]
\begin{center}
$%
\begin{array}{c@{\hspace{.1in}}cc}
\includegraphics[width=2.8in]{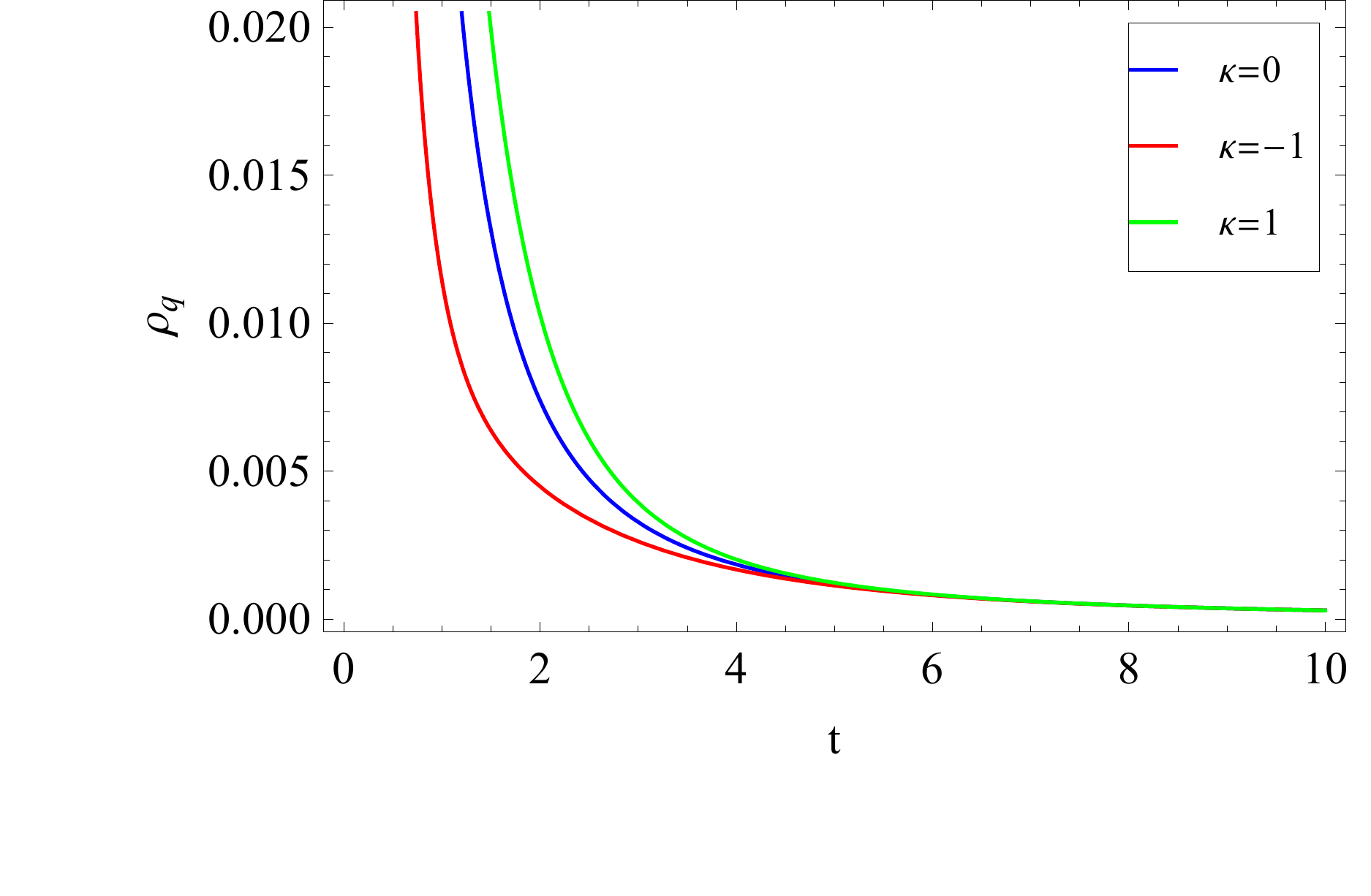} & 
\includegraphics[width=2.8in]{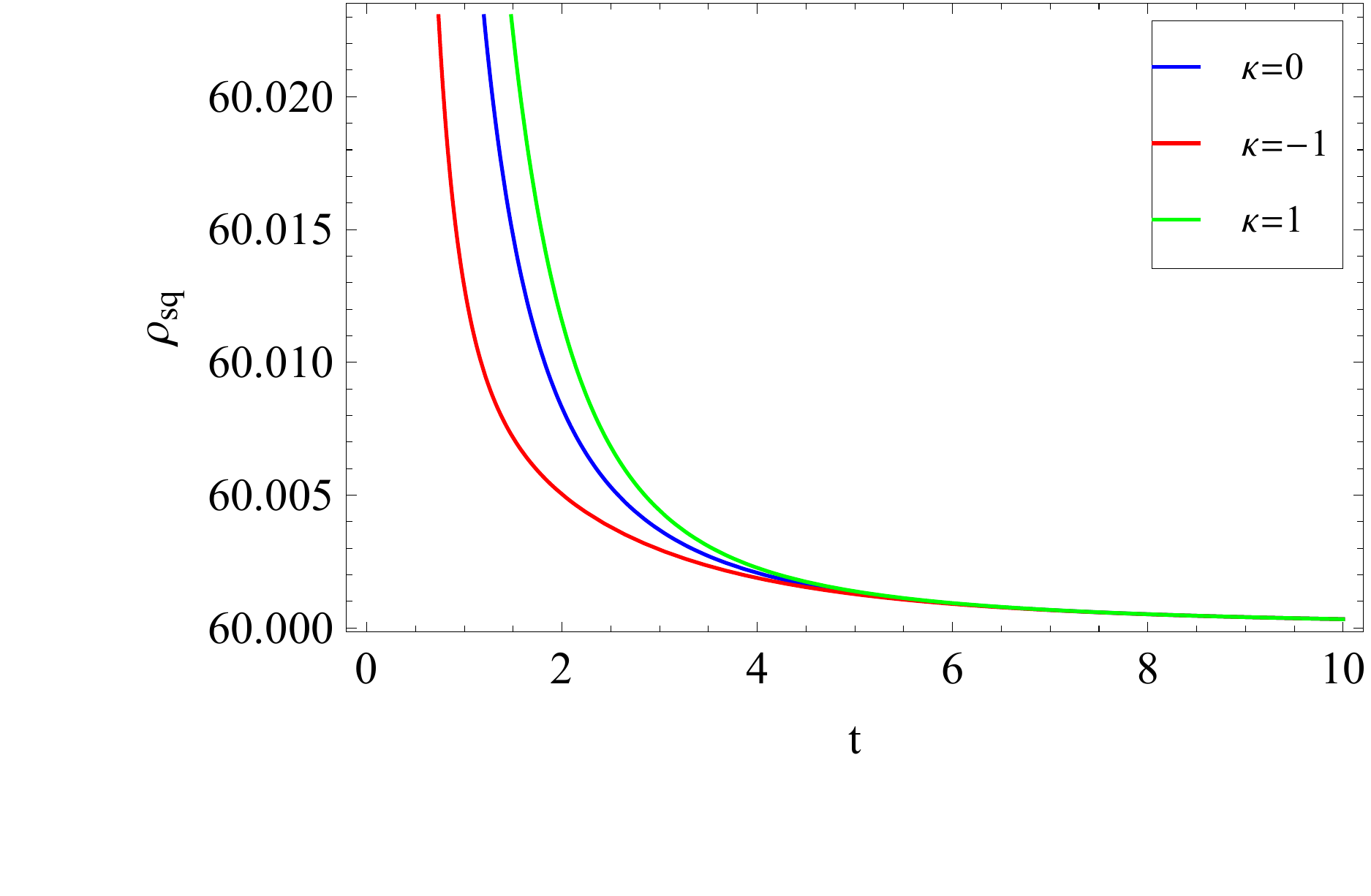}  \\ 
\mbox (a) & \mbox (b)%
\end{array}%
$%
\end{center}
\caption{\scriptsize (a) \textit{ The plot of $ \rho_{q} $  Vs. $ t $ for QM for Model-II }.\,\,\,  (b) \textit{ The plot of $ \rho_{sq} $  Vs. $ t $ for SQM for Model-II. }}
\end{figure}

\begin{figure}[tbh]
\begin{center}
$%
\begin{array}{c@{\hspace{.1in}}cc}
\includegraphics[width=2.8in]{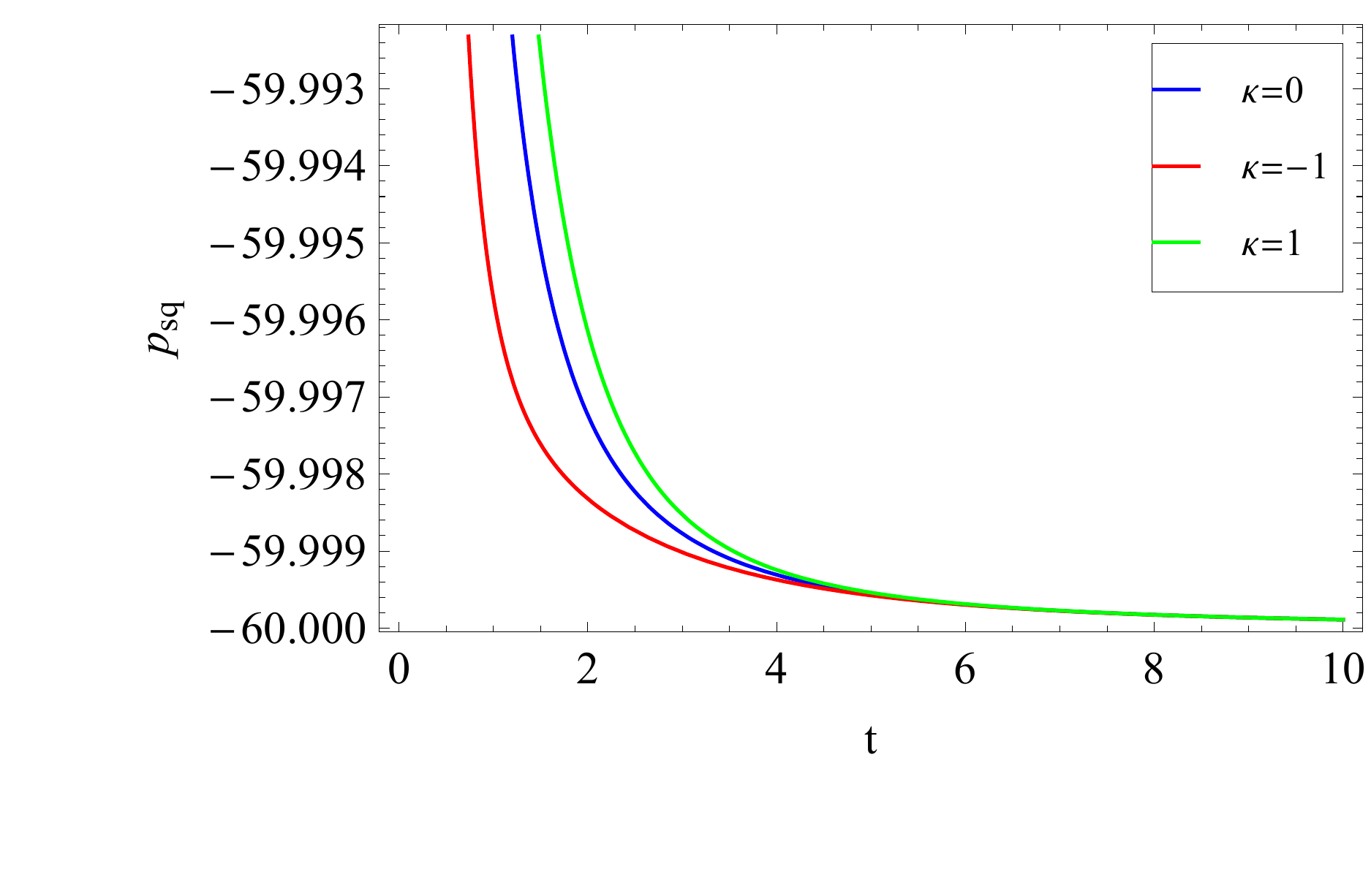} & 
\includegraphics[width=2.8in]{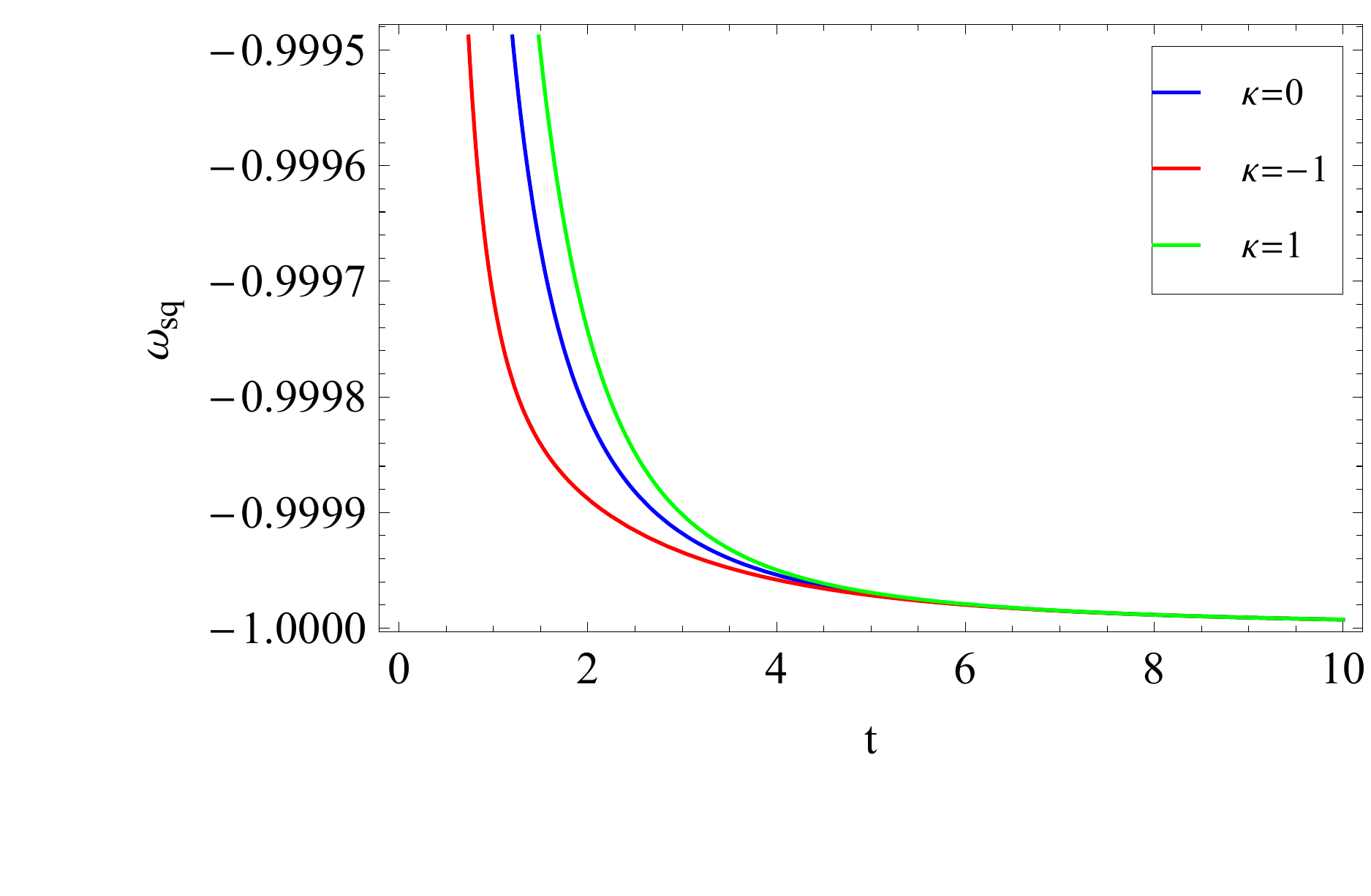}  \\ 
\mbox (a) & \mbox (b)%
\end{array}%
$%
\end{center}
\caption{\scriptsize (a) \textit{ The plot of $ p_{sq} $  Vs. $ t $ for SQM for Model-II }.\,\,\,  (b) \textit{ The plot of $ \omega_{sq} $  Vs. $ t $ for SQM for Model-II. }}
\end{figure}

\begin{figure}[tbh]
\begin{center}
$%
\begin{array}{c@{\hspace{.1in}}cc}
\includegraphics[width=2.8in]{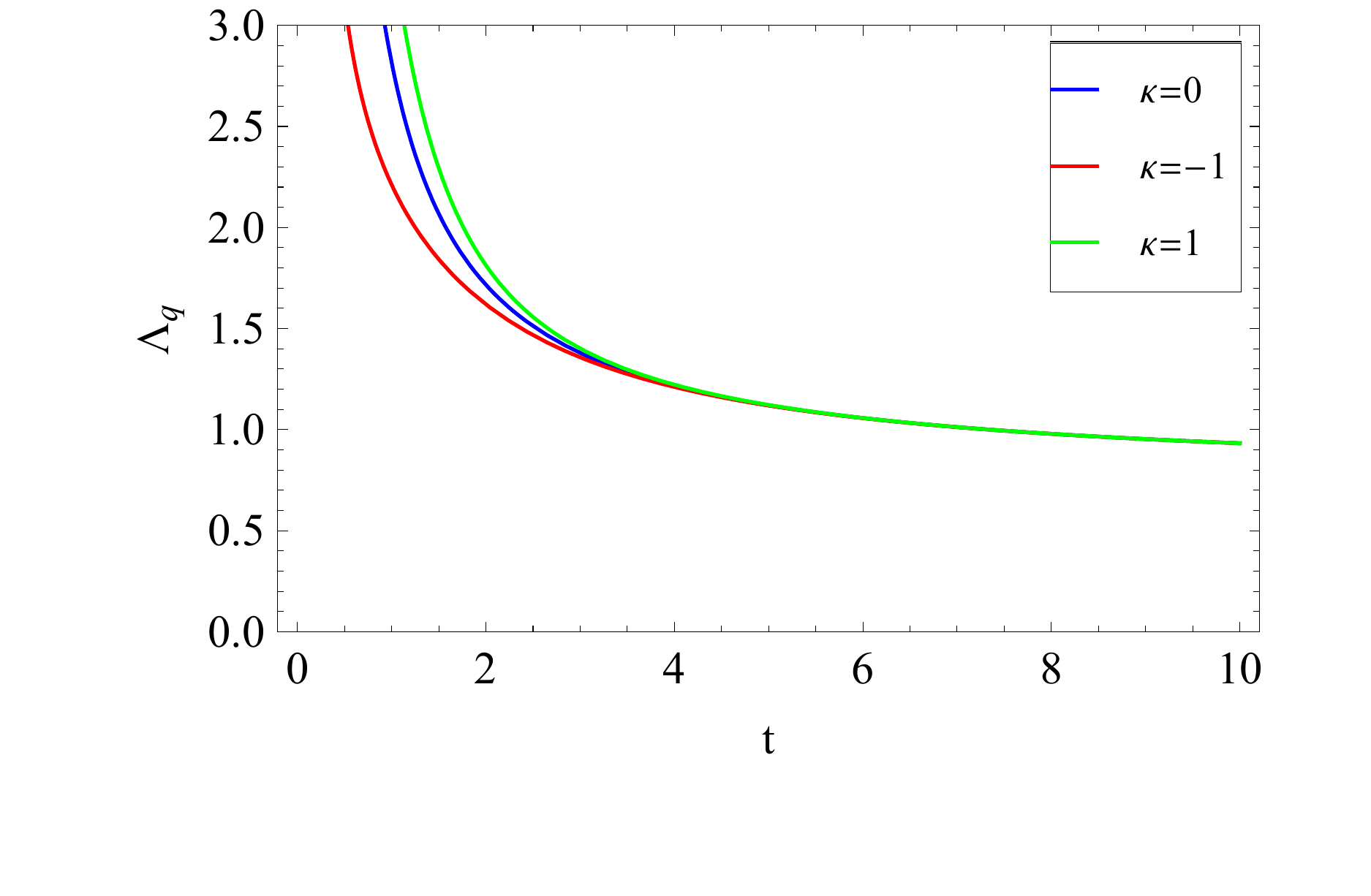} & 
\includegraphics[width=2.8in]{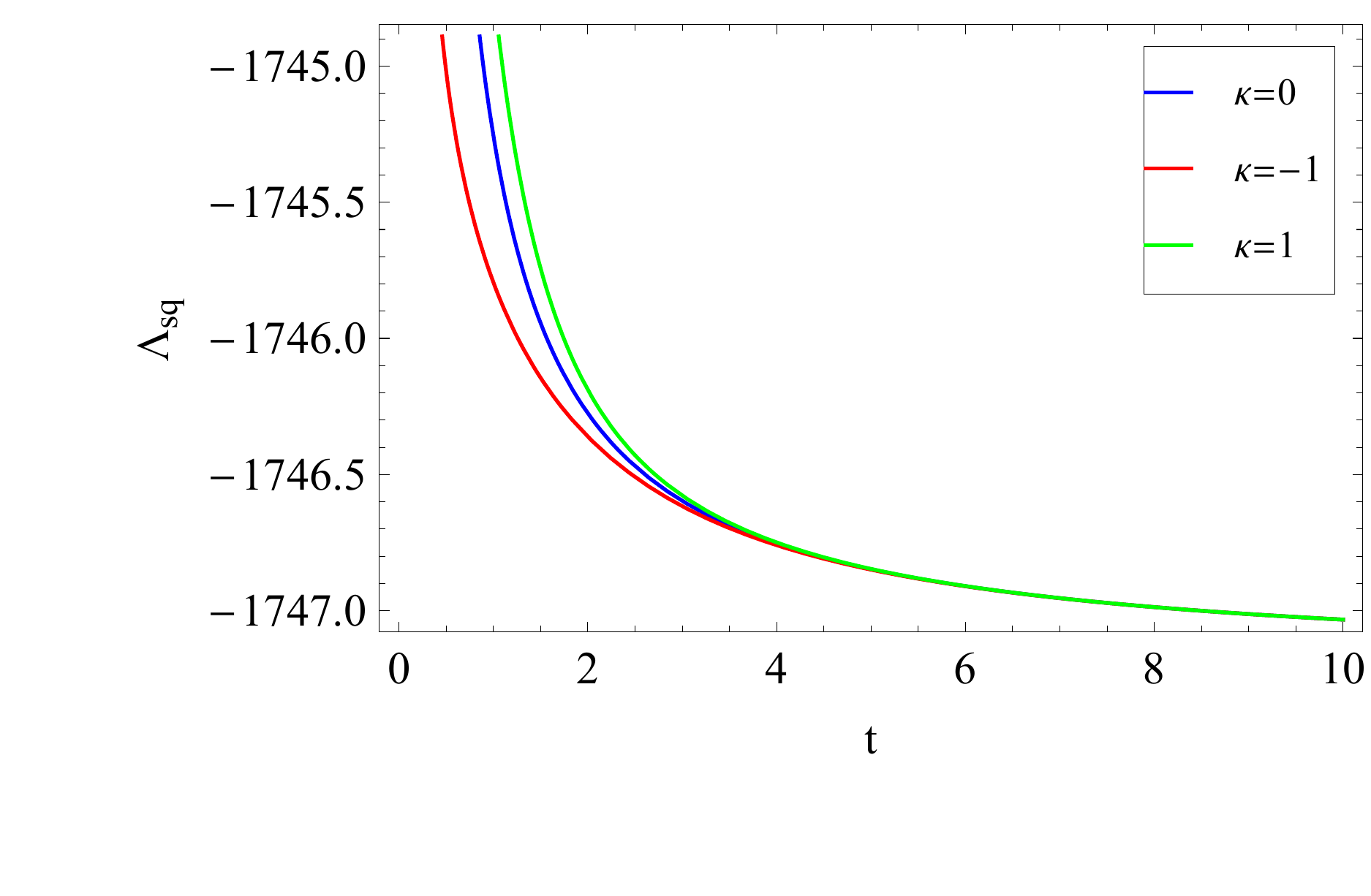}  \\ 
\mbox (a) & \mbox (b)%
\end{array}%
$%
\end{center}
\caption{\scriptsize (a) \textit{ The plot of $ \Lambda_{q} $  Vs. $ t $ for QM for Model-II }.\,\,\,  (b) \textit{ The plot of $ \Lambda_{sq} $  Vs. $ t $ for SQM for Model-II. }}
\end{figure}

Fig. 4(a) shows the profile of energy density $ \rho_q $ for QM  and SQM in Model II. Initially at the time of  evolution of the Universe, the amount of $ \rho_q $ is very large which steadily reduces with time and $ \rho_q \to 0 $ as $ t \to \infty $ in all the three different geometries of the Universe ($\kappa=0,-1,1$). The Universe has maximum value of energy density near $t=0$. Fig. 4(b) represents the behavior of energy density $\rho_{sq}$ for SQM in Model II. It has been observed readily that $\rho_{sq}$ behaves alike $\rho_q$ with only one diversity is that $\rho_{sq} \to B_c$ as $t \to \infty$.\\

Fig. 5 demonstrates the profile of pressure $p_{sq}$ and $\omega_{sq}$ for SQM. Fig. 5(a) describes the pressure profile $p_{sq}$ of Model II which is a decreasing function of time, remains negative throughout the evolution and $p_{sq}\rightarrow -B_{c}$ as $t\rightarrow \infty $. The negative pressure $ p_{sq} $ corresponds to the accelerating expansion of the Universe. Fig. 5(b) exhibits the behavior of EoS parameter $ \omega_{sq} $ for SQM in Model II. The EoS parameter $ \omega_{sq} $ belongs to the quintessence domain, decreases with time and finally approaches to phantom divide line \textit{i.e.} $\omega_{sq} \to -1$ in late time which is affirmative with the current observations. Thus, in this case, we see that SQM behaves like dark energy.\\  

Fig. 6(a) illustrates the behavior of cosmological constant $\Lambda_q$ for QM. In Model II, $\Lambda_q$ is a decreasing function with time and $\Lambda_q \to 1$ as $t \to \infty$. Fig. 6(b) highlights the action of cosmological constant $\Lambda_{sq}$. In Model II, $\Lambda_{sq}$ decreases with time remains negative throughout.\\ 

\subsection{ESU Model}

\qquad If we use $a(t)^2=1$ in Eq. (\ref{3}) we transform FLRW universe to Einstein Static Universe (ESU), also we get new metric as follows
\begin{equation} \label{eq35}
ds^2=-dt^2+\frac{dr^2}{1-kr^2}+r^2(d\theta^2+\sin^2\theta d\phi^2)
\end{equation}
Using Einstein Static Universe model with Eqs. (\ref{7})-(\ref{9}) and Eqs. (\ref{10})-(\ref{12}), we get the values of cosmic pressure, density, magnetic field and cosmological  constant solutions for QM and SQM distributions in $f(R,T)$ gravity and GR as follows\\
\begin{scriptsize}
\begin{center}
\textbf{Table 3.} The behavior of physical parameters for ESU in $f(R,T)$ theory \\[10pt]
\begin{tabular}{|c|c|c|}
	\hline 
ESU	in $f(R,T)$& QM & SQM \\ 
	\hline 
$\rho$	& $\dfrac{\kappa}{(4\pi+\lambda)(1+\omega)}$ &  $\dfrac{3\kappa}{4(4\pi+\lambda)}+B_c $ \\
	\hline 
$p$	& $\dfrac{\kappa \omega}{(4\pi+\lambda)(1+\omega)}$ & $\dfrac{\kappa}{4(4\pi+\lambda)}-B_c $  \\ 
\hline 
$\Lambda$	& $ \dfrac{4\kappa(3\pi+\lambda)\omega+4\pi }{(4\pi + \lambda)(1+\omega)} $  & $ \dfrac{(6\pi+\lambda)\kappa}{(4\pi+\lambda)}-4B_c(2\pi+\lambda) $ \\ 
\hline 
$h^2$	& $0$  & $0$ \\ 
	\hline 
\end{tabular} 
\end{center}

\begin{center}
\textbf{Table 4.} The behavior of physical parameters for ESU in GR \\[10pt]
\begin{tabular}{|c|c|c|}
	\hline 
	ESU	in GR& QM & SQM \\
	\hline 
	$\rho$	& $\dfrac{\kappa}{4\pi(1+\omega)}$ &  $\dfrac{3\kappa}{16\pi}+B_c $ \\
	\hline
	$p$	& $\dfrac{\kappa \omega}{4\pi(1+\omega)}$ & $\dfrac{\kappa}{16\pi}-B_c $  \\ 
	\hline 
	$\Lambda$	& $ \dfrac{3\kappa \omega+1 }{1+\omega} $  & $ \dfrac{3\kappa}{2}-8\pi B_c $ \\ 
	\hline 
	$h^2$	& $0$  & $0$ \\ 
	\hline 
\end{tabular} 
\end{center}
\end{scriptsize}

\section{Interpretation and validation of models}
\subsection{Energy conditions}

\qquad The dynamic model, which is independent of constraints on the kinematics of the Universe can be further obtained from the energy conditions. These conditions impose restriction on the energy-momentum tensor $T_{\mu\nu} $, and can be transformed into inequalities restricting the possible value of the pressure and density of the fluid \cite{haw,scar,vis,san,zhang}. The energy conditions in terms of pressure and energy density of the fluid take
the form as
\begin{center}
\indent $NEC $: $\,\,\,\,\,\rho+p\geq0 $,\\
\indent $WEC $: $\,\,\,\,\,\,\,\,\,\,\,\rho\geq0$, $\,\,\,
\,\,\,\,\,\,\,\,\,\rho+p\geq0 $,\\[0pt]
\indent $SEC $: $\,\,\,\rho+3p\geq0$, $\,\,\,\,\,\,\,\,\,\,\,\,\rho+p\geq0 $,\\
\indent $DEC $: $\,\,\,\,\,\,\,\,\,\,\,\,\,\rho\geq0$, $\,\,\,\,\,\,\,\,\,
-\rho\leq p\leq \rho $,\\[0pt]
\end{center}

\noindent where NEC, WEC, SEC, and DEC represent the Null, Weak, Strong and Dominant energy conditions respectively. The energy conditions are independent of EoS parameter for matter filled Universe. These conditions provide a very simple constraints on the nature of the pressure and energy density of the fluid. Therefore the energy conditions provide some aspects for explaining the evolution of the Universe depends on some general principles.
{\scriptsize 
\begin{table*}[tbh]
\begin{center}
{\scriptsize \textbf{Table 5.} The status of energy conditions for QM and SQM for Model I and Model II, 
where $\oplus $ and $ \ominus $ \\
indicate the different energy conditions satisfied and unsatisfied respectively. \\[10pt]
\begin{tabular}{|c|cccc|}
\hline
Matter & NEC & WEC & SEC & DEC \\ \hline
QM & $\oplus $ & $\oplus $ & $\oplus $ & $\oplus $ \\ \hline
SQM & $\oplus $ & $\oplus $ & $\ominus$ & $\oplus $ \\ \hline
\end{tabular}
\label{table:nonlin}  }
\end{center}
\end{table*}
} 
In Model I, we have discussed energy conditions for flat and closed geometry of the Universe while in Model II, energy conditions for all the flat, open and closed geometry of the Universe have been discussed. From Table 3, it can be apparently seen that all the energy conditions are satisfied positively for QM. For SQM, the NEC, WEC and DEC hold good but SEC fails to satisfy which indicates that the model undergoes the accelerated expansion of the Universe (see Fig. 2b, 5b). 

\subsection{f(R,T) function}

\subsubsection{Model I}

\qquad The Ricci scalar $ R $ and trace $ T=\rho-3p$ of the stress energy momentum tensor for the QM and SQM of Model I is given by 
\begin{eqnarray}  \label{41}
R = \frac{6(-\alpha+(\beta-2)t^2)}{(\alpha+\beta t^2)^2}-\frac{6\kappa}{(\alpha+\beta t^2)^{\frac{1}{\beta}}},
\end{eqnarray}

\begin{eqnarray}  \label{42}
T_q = \frac{-(\alpha-\beta t^2)(3\omega -1)}{(4\pi+\lambda)(1+\omega)(%
\alpha+\beta t^2)^2}+\frac{(1-3\omega)\kappa}{(4\pi+\lambda)(1+\omega)(\alpha+\beta t^2)^{\frac{1}{\beta}}},
\end{eqnarray}
\begin{eqnarray}  \label{43}
T_{sq} = 4B_c.
\end{eqnarray}
From the above Eqs. (\ref{41})-(\ref{43}) one can find the function $ f(R,T)=R+2f(T)$ for QM and SQM of the form 
\begin{eqnarray}  \label{44}
f_q(R,T) =\frac{6(-\alpha+(\beta-2)t^2)}{(\alpha+\beta t^2)^2}-\frac{6\kappa}{(\alpha+\beta t^2)^{\frac{1}{\beta}}}- \frac{2\lambda (\alpha-\beta t^2)(3\omega -1)}{(4\pi+\lambda)(1+\omega)(\alpha+\beta t^2)^2}\nonumber \\+ \frac{2\lambda(1-3\omega)\kappa}{(4\pi+\lambda)(1+\omega)(\alpha+\beta t^2)^{\frac{1}{\beta}}},
\end{eqnarray}
\begin{eqnarray}  \label{45}
f_{sq}(R,T) = \frac{6(-\alpha+(\beta-2)t^2)}{(\alpha+\beta t^2)^2}-\frac{6\kappa}{(\alpha+\beta t^2)^{\frac{1}{\beta}}}+8\lambda B_c.
\end{eqnarray}

\subsubsection{Model II}

\qquad The Ricci scalar $R$ and trace $T=\rho-3p$ of the stress energy momentum tensor for the QM and SQM of Model II is given by 
\begin{eqnarray}  \label{46}
R =\frac{6(k_2-2(k_1 t+k_2)^2)}{t^2}-\frac{6\kappa}{t^{2k_2}e^{2k_1 t} },
\end{eqnarray}
\begin{eqnarray}  \label{47}
T_q = \frac{(1-3\omega)k_2}{(4\pi+\lambda)(1+\omega)t^2}+\frac{(1-3\omega)\kappa}{(4\pi+\lambda)(1+\omega)t^{2k_2}e^{2k_1 t}  },
\end{eqnarray}
\begin{eqnarray}  \label{48}
T_{sq} = 4B_c.
\end{eqnarray}
From the above Eqs. (\ref{46})-(\ref{48}) one can find the function $ f(R,T)=R+2f(T)$ for QM and SQM of the form 
\begin{equation} \label{49}
f_q(R,T)=\frac{6(k_2-2(k_1 t+k_2)^2)}{t^2}-\frac{6\kappa}{t^{2k_2}e^{2k_1 t} }+\frac{2\lambda(1-3\omega)k_2}{(4\pi+\lambda)(1+\omega)t^2}+\frac{2\lambda(1-3\omega)\kappa}{(4\pi+\lambda)(1+\omega)t^{2k_2}e^{2k_1 t}}
\end{equation}%
\begin{equation} \label{50}
f_{sq}(R,T)=\frac{6(k_2-2(k_1 t+k_2)^2)}{t^2}-\frac{6\kappa}{t^{2k_2}e^{2k_1 t} }+8\lambda B_{c}.
\end{equation}

\subsection{Scalar field correspondence for SQM}

\qquad Looking at the structure of the EoS parameter for SQM in both the models, we can interpret that the matter field of SQM as some kind of exotic matter field with dominant potential that can produce some antigravitational effect with negative pressure making the EoS parameter $w$ negative. So, we can consider this matter field as a scalar field as usual for other scalar field models (\textit{e.g.} quientessence, phantom \textit{etc.}). Also in a paper by Rahaman \textit{et al.} \cite{far}, we have observed that MIT Bag model of quark matter shows curious characteristics of quark matter. Bag model of quark matter can behave like both dark matter and dark energy depending upon the pressure term $ p $ to be positive and negative. By this reason we hypothesize that if scalar field $ \phi $ is the source of the exotic matter field having a self interacting potential $ V(\phi ) $, then by adopting Barrow's scheme \cite{barr}, one can write energy density $\rho_{\phi}$ and pressure $p_{\phi}$ for the scalar field as, 
\begin{equation}\label{51}
\rho _{\phi }=\frac{1}{2}\dot{\phi}^{2}+V(\phi )=\rho ,  
\end{equation}%
\begin{equation}\label{52}
p_{\phi }=\frac{1}{2}\dot{\phi}^{2}-V(\phi )=p.  
\end{equation}%
Using Eqs. (\ref{28}) and (\ref{29}), we have the value of $ V(\phi) $ and $\phi $ for SQM in case of Model I for flat Universe ($\kappa=0$) as:

\begin{equation} \label{53}
\phi _{sq}= t\,ln\,t\sqrt{\frac{2-\beta}{(4\pi +\lambda )(2-3\beta )^{2}t^{2}}}+\phi _{0},
\end{equation}%
where $\phi_{0}$ is the constant of integration.

\begin{equation} \label{54}
V(\phi )_{sq}= \frac{-\alpha +\beta t^{2}}{2(8\pi +2\lambda )(\alpha +\beta t^{2})^{2}}+B_{c}.
\end{equation}
Similarly, using Eqs. (\ref{36}) and (\ref{37}), we have the value of $ V(\phi) $ and $\phi $ for SQM in case of Model II for flat Universe ($\kappa=0$) as

\begin{equation}\label{55}
\phi _{sq}= t\,ln\,t\sqrt{\frac{k_{2}}{(4\pi +\lambda )t^{2}}}+\phi _{1},
\end{equation}
where $\phi_{1}$ is the constant of integration.

\begin{equation}\label{56}
V(\phi )_{sq}= \frac{k_{2}}{2(8\pi +2\lambda )t^{2}}+B_{c}.
\end{equation}\\
\begin{figure}[tbh]
\begin{center}
$%
\begin{array}{c@{\hspace{.1in}}cc}
\includegraphics[width=3in]{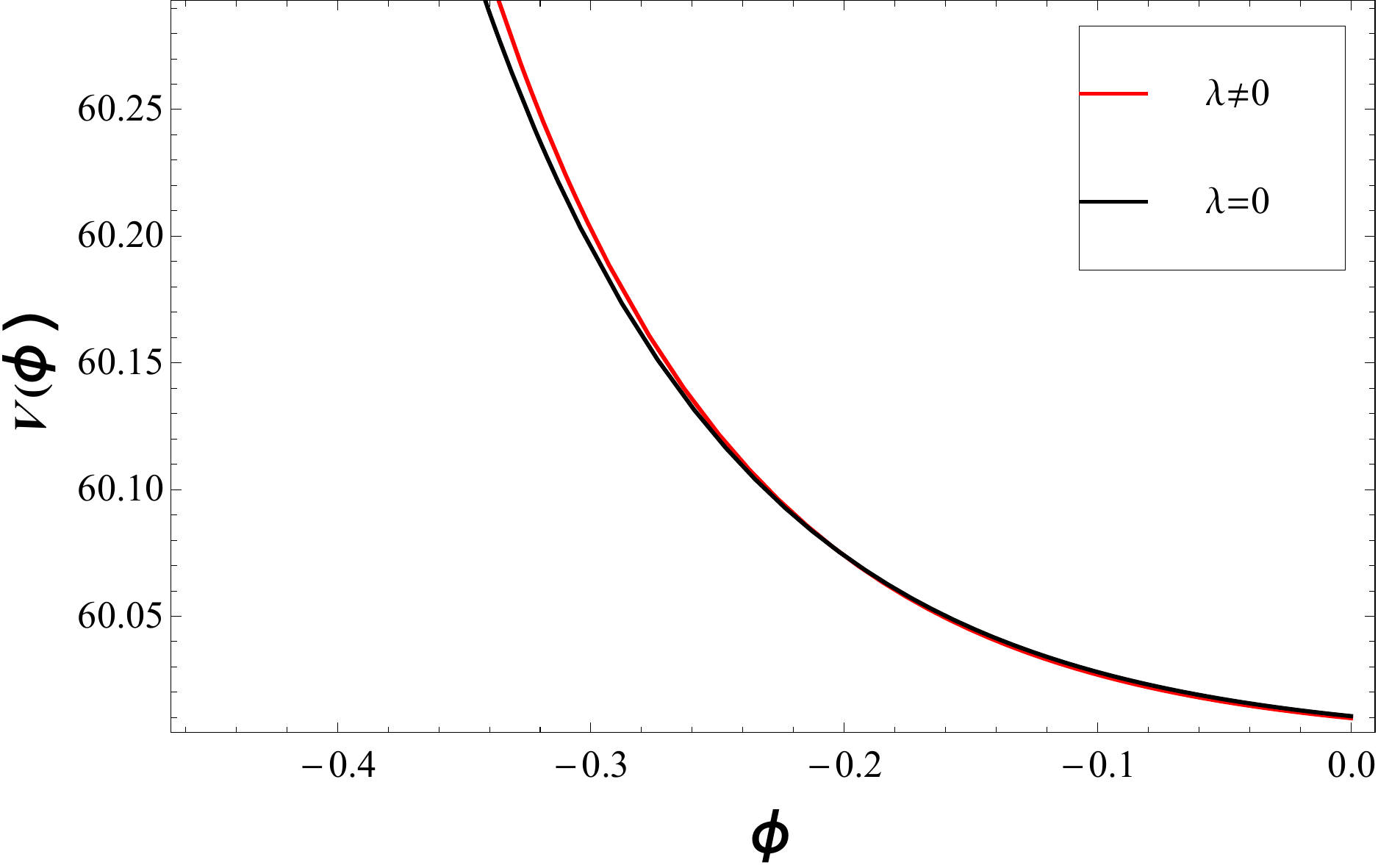} & %
\includegraphics[width=3in]{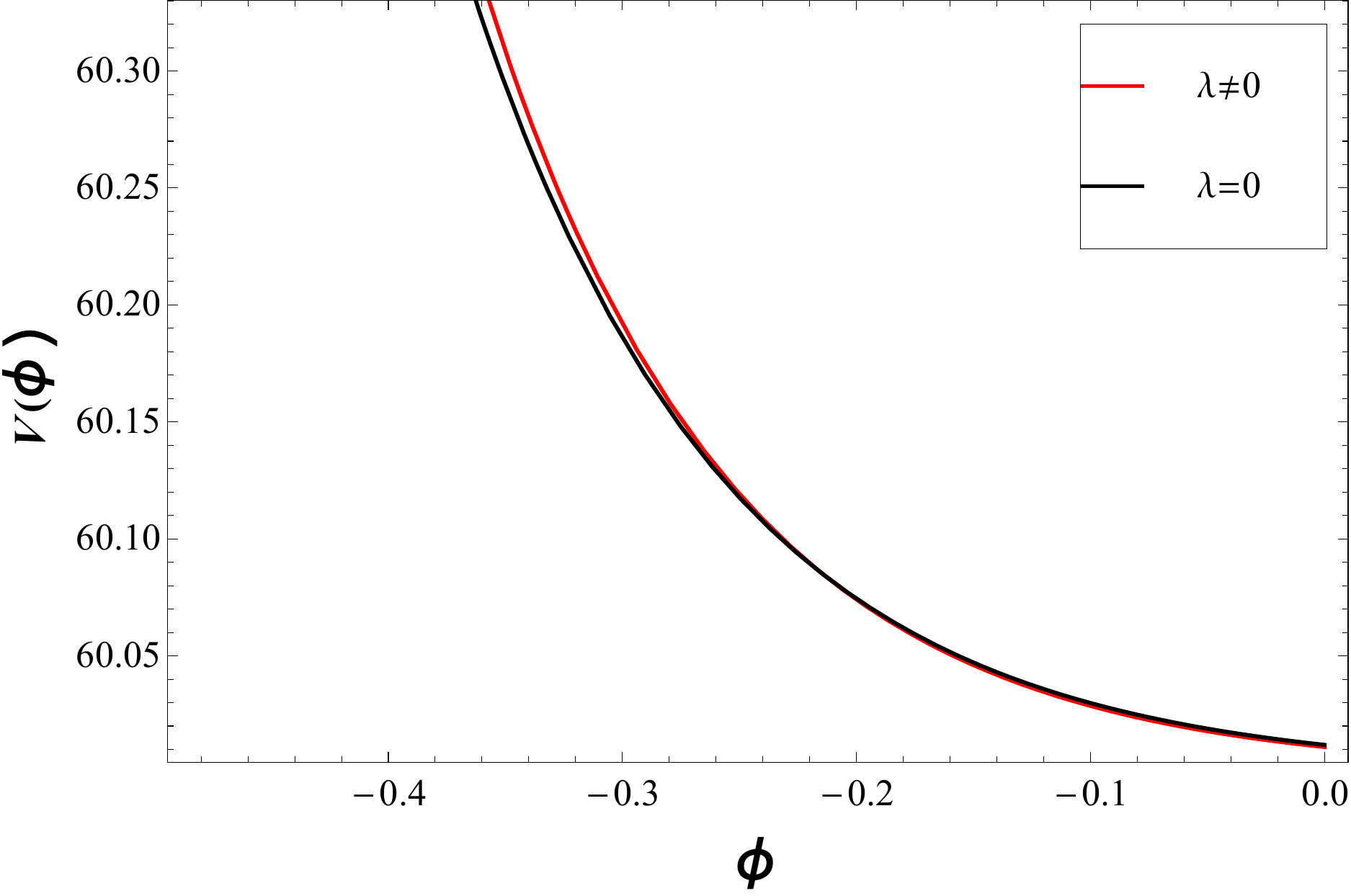} \\
\mbox (a) & \mbox (b)%
\end{array}%
$%
\end{center}
\caption{\scriptsize (a) \textit{The plot of }$ V(\protect\phi) $\textit{\ Vs.} $ \phi $
\textit{\ for SQM for Model I }. (b) \textit{The plot of } $ V(\phi )$ \textit{\ Vs. } $ \phi $ \textit{\ for SQM for Model II.}}
\end{figure}

In this section, we have discussed the scalar field $ \phi $ and scalar potential $ V(\phi) $ for SQM in case of $ f(R,T) $ gravity and GR for flat Universe only. The graphical representations of scalar potential $ V(\protect\phi) $ for SQM in case of $ f(R,T) $ gravity and GR  is shown in Fig. 7a with the values $ \mathit{\protect \lambda =1} $, $ \beta=1.1 $ and $ \mathit{B_{c}=60} $ for Model I, and in Fig. 7b with the values $\mathit{\protect \lambda =1 }$, $ k_{1}=0.5 $, $ k_2=0.6 $ and $\mathit{B_{c}=60}$ for Model II. The scalar potential $ V(\phi) $ is monotonically decreases with scalar field $ \phi $ in case of $ f(R,T) $ gravity and GR for Model I and II. The flat plateau of the potential energy $ V(\phi) $ indicates the accelerating expansion of the Universe in case of  $ f(R,T) $ gravity and GR for both models. 

\subsection{Velocity of sound}

\qquad In order to determine the stability of any cosmological model, one can examine a stringent condition \textit{i.e.} velocity of sound $ C_{s}^{2} <1 $ . This lead to avoid unwanted fluctuations in power spectrum of matter. In this study, velocity of sound, $C_{s}^{2}= \frac{dp}{d\rho}$ in Model I and Model II behaves alike for both QM and SQM respectively. \\
\begin{itemize}
\item For quark matter, $ C_{s}^{2}= \frac{dp}{d\rho}= \omega $, where $ 0\leq \omega \leq 1 $ for Model I and II,
\item For strange quark matter, $C_{s}^{2}= \frac{dp}{d\rho}= \frac{1}{3}$, which is less than $ 1 $ for Model I and II.\\
\end{itemize}
Hence, we see that both of our models for QM as well as SQM are stable. 

\section{Discussion and conclusion}

\qquad In this paper, we have explored the FLRW model of the Universe based on f(R,T) gravity with magnetized QM as well as SQM as the source of matter. The physical parameters \textit{i.e.} energy density, isotropic pressure, cosmological constant \textit{etc.} for QM and SQM have been investigated by considering two different parametrizations \textit{i.e.} deceleration parameter $ q $, and the expansion factor $ a $, which have been considered as Model I and II respectively. The energy conditions of the models have been discussed. The behavior of $ f(R,T) $ function, the scalar field correspondence of SQM, and the stability of the models have been studied. A detailed observations have been done for Model I and Model II as follows:

\begin{itemize}
\item Model I exhibits phase transition at $ t_{ph}=\sqrt{\frac{\alpha }{\beta -1}} $, and shows bouncing behavior hence it is free from initial singularity. Dynamics of the Universe depends on the model parameters $ \alpha $ and $\beta $. The model exhibits decelerated expansion with the constraints $ \alpha <0 $ and $ \beta >1 $ or $ 0<\alpha <(\beta -1)t^{2} $ and $ \beta >1 $, and accelerated expansion with $ \alpha >(\beta -1)t^{2} $ and $ \beta >1 $.   Model II also exhibits a phase transition from early deceleration to late time acceleration at $ t_{ph}=\frac{-k_{2}}{k_{1}}+\frac{\sqrt{k_{2}}}{k_{1}} $, which follows the standard big-bang scenario.  The choice of these two parametrization provides a comparative study of two different models. The model I solve the initial singularity problem where as the model II describes the late time acceleration well.

\item We have observed that in our isotropic Universe, magnetic flux is ineffective \textit{i.e.} $ h^2=0 $ in both models for QM and SQM. Also, using constant deceleration parameter Akta{\c s} and Ayg\"{u}n \cite{can} investigated magnetized SQM distributions in $f(R,T)$ gravity for first and second models of Harko et al. \cite{har}. They have obtained $h^2=0$ for these models. Our results agree with \cite{can} in $f(R,T)$ gravity and GR theory. But for anisotropic case, $ h^2 $ is time dependent, non-vanishing and $ h^2 \to 0 $ as $ t \to \infty $ \cite{sah2}.

\item In case of Model I, the physical parameters for QM and SQM can be studied through various plots for flat and closed Universe. The energy densities for QM and SQM are monotonically decreasing with time and $\rho_q \to 0$ as $t \to \infty$ and $\rho_{sq} \to B_c $ as $t \to \infty$ (see Fig. 1a, 1b). The pressure $ p_{sq} $ for SQM is monotonically decreasing with time, remains negative throughout the evolution of the Universe, and $ p_{sq}\rightarrow -B_{c} $ as $ t\rightarrow \infty $. The negative pressure $ p_{sq} $ corresponds to the accelerating expansion of the Universe (see Fig. 2a). The EoS parameter $\omega_{sq} $ for SQM reside in the quintessence domain, decreases monotonically with time $ t $, and finally approaches to phantom divide line at late time \textit{i.e.} $ \omega_{sq} \to -1 $ as $ t \to \infty $, which confirms the current observations (see Fig. 2b). Model I shows physically realistic model for flat and closed Universe only.

\item In case of Model II, the energy density for QM and SQM are monotonically decreasing with time and $\rho_q \to 0 $ as $ t \to \infty $ and $ \rho_{sq} \to B_c $ as $ t \to \infty $ (see Fig. 4a, 4b) for flat, closed and open Universe. The energy density is maximum at $ t\simeq0 $. The pressure $ p_{sq} $ for SQM monotonically decreases with time, remains negative throughout the evolution of the Universe, and $ p_{sq}\rightarrow -B_{c} $ as $ t\rightarrow \infty $. The negative pressure $ p_{sq} $ corresponds to the accelerating expansion of the Universe (see Fig. 5a). The EoS parameter $ \omega_{sq} $ for SQM decreases monotonically with time $ t $, exists in the quintessence domain and finally approaches to phantom divide line at late time \textit{i.e.} $ \omega_{sq} \to -1 $ as $ t \to \infty $, which is consistent with the current observations (see Fig. 5b). Thus, in this case, we see that SQM behaves like dark energy. Model II represents the physically acceptable model for flat, closed and open Universe.

\item The behavior of cosmological parameter for QM and SQM in Model I is shown in Fig 3 for flat and closed FRW models. In Model I, $ \Lambda_q $ and $ \Lambda_{sq} $ are monotonically decreasing functions of time, and $ \Lambda_q \to 0 $ as $ t \to \infty $. The cosmological parameter $ \Lambda_{sq} $ starts with negative value and  tends to a negative finite quantity as $ t\to\infty $ \textit{i.e.} it remains negative throughout the evolution of the Universe. 

\item Fig. 6(a) and 6(b) illustrate the behavior of cosmological parameters $ \Lambda_q $  and $\Lambda_{sq}$ for QM and SQM respectively for flat, closed and open Universe in Model II. In this model, both $ \Lambda_q $  and $\Lambda_{sq} $ are monotonically decreasing function of time and $ \Lambda_q \to 1 $ and  $ \Lambda_{sq} \to a\,\, finite\,\, negative\,\, quantity $ as $ t \to \infty $. Therefore, we observe that $\Lambda_{sq}$ decreases with time and remains negative throughout the evolution of the Universe.
 
\item In subsection 4.3, we have obtained Einstein Static Universe (ESU) for FLRW model by using $a(t)^2=1$ in Eq. (\ref{3}) and studied physical features of the obtained model (see Table 3,4).  

\item In section 5, all the energy conditions NEC, WEC, SEC, and DEC are satisfied for QM in Model I and II. For SQM in both Model I and II, SEC fails to satisfy which agrees with present observation of accelerated Universe as recent acceleration of Universe demands an exotic matter satisfying $1+3\omega < 0$ (see Table 5).

\item The structure of the EoS parameter for SQM in Model I and II motivated us to look for its scalar field correspondence for which we have obtained the potential energy $V(\phi )$ and scalar field $ \phi $ in $ f(R,T) $ gravity as well as in GR ($ \lambda=0 $) for flat Universe only whose behavior is shown in Fig. 7a, 7b. The flat plateau of the potential energy $ V(\phi) $ indicates the accelerating expansion of the Universe in case of  $ f(R,T) $ gravity and GR for both models, which is in good agreement with current observations.

\item Although we see traces of magnetic fields all over our Universe, they are often very weak and difficult to characterize. According to observational and theoretical  studies the Milky Way is host to a magnetic field on the order of $10^{-6} G$. Also the strengths of magnetic field approximately $10^{-9} G$ maybe even smaller in the early Universe. But we exactly don’t know their role in how our Universe has evolved \cite{Marinacci}. According to these results, our magnetic field value is consistent for the homogeneous and isotropic FRW Universe model in $f(R,T)$ gravity. However, for $t \to \infty$ we get  $\rho_{sq} \to B_c $, $p_{sq}\to -B_c$ also $\omega_{sq} \to -1$ and small negative cosmological parameter as $\Lambda=-4(2\pi+\lambda)B_c$ in $f(R,T)$ gravity and $\Lambda=-8\pi B_c$ in GR theory for Model I and II. These solutions correspond to the accelerating expansion of the Universe. So, using the different dynamics of the Universe, the obtained dark energy results for SQM are consistent with observations of the type Ia Supernovae \cite{per},\cite{rie}.
\end{itemize}
 
\noindent \textbf{Acknowledgements} The authors express their thanks to Prof. M. Sami, Prof. S. G. Ghosh, Centre for Theoretical Physics, Jamia Millia Islamia, New Delhi, India and Prof. J. P. Saini, Director, NSIT, New Delhi, India for some fruitful discussions and providing necessary facilities to complete the work. Authors also express their thanks to referee for his valuable comments.\\

\end{document}